\documentclass[preprint]{ptephy_v1}%%%%%% to generate preprint number

\preprintnumber{KYUSHU-HET-231, KOBE-TH-21-02} %%% %%% Insert preprint number here

\usepackage{amsmath} %for dealing with mathematics,
\usepackage{amsthm} %for dealing with theorem environments,
\usepackage{url} %It provides better support for handling and breaking URLs.

\usepackage{stmaryrd}

\DeclareMathOperator{\Tr}{Tr}
\newcommand{\Slash}[1]{{\ooalign{\hfil/\hfil\crcr$#1$}}}
\numberwithin{equation}{section}
\allowdisplaybreaks

\begin{document}

\title{Manifestly gauge invariant exact renormalization group\\
for quantum electrodynamics}

%%%% To generate auto affiliation numbers please use \author{}\affil{} command

\author{Yuki Miyakawa}
\affil{Department of Physics, Kyushu University, 744 Motooka, Nishi-ku,
Fukuoka 819-0395, Japan}

\author{Hidenori Sonoda}
\affil{Physics Department, Kobe University, Kobe 657-8501, Japan}

\author[1]{Hiroshi Suzuki}

%% \author{Insert second author name here}
%% \affil{Insert second author address here}

%% \author{Insert third author name here}
%% \author[3]{Insert fourth author name here} %%% Use optional bracket [3] to change the respective address
%% \affil{Insert third author address here}

%% \author{Insert last author name here\thanks{These authors contributed equally to this work}}
%% \affil{Insert last author address here}

%%% To include the collaborator name... Please use the command "\collaborator"
%%% For example: \collaborator{ATLAS Collaboration}

\begin{abstract}%
We formulate quantum electrodynamics on the basis of gauge (or BRST) covariant
diffusion equations of fields. This is a particular example of the gradient
flow exact renormalization group (GFERG). The resulting Wilson action fulfills
a simple gauge Ward--Takahashi identity. We solve the GFERG equation around the
Gaussian fixed point to the second order in gauge coupling and obtain the
1-loop beta function and anomalous dimensions. The anomalous dimension of the
electron field coincides with that of the fermion field diffused by a gauge
covariant flow equation of L\"uscher.
\end{abstract}

\subjectindex{B05, B32}

\maketitle

\section{Introduction}
\label{sec:1}
A Wilson action is a functional of field variables with a finite momentum
cutoff, say~$\Lambda$~\cite{Wilson:1973jj}. If the underlying theory is a
continuum limit, the theory is defined to all momentum scales. We obtain the
interaction vertices of the Wilson action~$S_\Lambda$ by integrating out the
fields with momenta larger than~$\Lambda$. It is then natural to expect that
only the correlations of the fields with momenta smaller than~$\Lambda$ are
kept, but those with momenta larger than~$\Lambda$ are lost from~$S_\Lambda$.
In the exact renormalization group (ERG) formalism~\cite{Wilson:1973jj}, in
which a sharp momentum cutoff is replaced by a smooth function of momentum,
this is not the case: we can still reconstruct the full correlation functions
using the Wilson action. This makes gauge invariance compatible with a momentum
cutoff. This viewpoint was first adopted for QED in~Ref.~\cite{Sonoda:2007dj}.
A general framework for constructing non-abelian gauge theories along this line
was given in~Ref.~\cite{Igarashi:2009tj}.

The realization of gauge invariance with a Wilson action has a long history
starting in the 1980's. The early works in the 1990's such
as~Refs.~\cite{Becchi:1996an,Ellwanger:1994iz,Bonini:1994dz,Bonini:1994kp,%
Reuter:1993kw,Reuter:1994sg} established the possibility of constructing gauge
theories in the ERG formalism. (Ref.~\cite{Ellwanger:1994iz} gives references
to the earlier works from the 1980's.)  What is common in the realization of
gauge invariance in the ERG formalism is that the gauge invariance is not what
one expects naturally. For~$\Lambda>0$, the gauge transformation is modified so
that the Jacobian is non-vanishing, and the resulting expression of gauge
invariance is by no means manifest. This has been an obstacle for any
calculation of the Wilson action beyond perturbation theory, since it is
difficult to truncate the action keeping the non-manifest gauge invariance.

The original formulation of ERG is based on the diffusion of the
fields~\cite{Wilson:1973jj}. Recently a proposal was made that we may be able
to construct a manifestly gauge invariant Wilson action by replacing the
diffusion equation by a gauge invariant diffusion
equation~\cite{Sonoda:2020vut}. This was inspired by the gauge invariant
diffusion that generates a gradient flow of gauge fields, first discussed
for lattice gauge theory in~Refs.~\cite{Narayanan:2006rf,Luscher:2009eq,%
Luscher:2010iy} and then by L\"uscher and Weisz~\cite{Luscher:2011bx} for
perturbative non-abelian gauge theory. We call this new type of ERG by the
gradient flow exact renormalization group (GFERG). The present paper is a
sequel to~Ref.~\cite{Miyakawa:2021hcx} where GFERG for fermions is discussed.

The paper is organized as follows. In~Sect.~\ref{sec:2} we first review the
relation between a diffusion equation and the exact renormalization group
transformation using a generic real scalar theory. We follow the discussions
given in~Ref.~\cite{Sonoda:2019ibh}; see also Ref.~\cite{Matsumoto:2020lha}. We
then introduce a particular set of diffusion equations for QED that is
consistent with the BRST invariance of the theory. We base our construction of
GFERG on these diffusion equations. In~Sect.~\ref{sec:3} we construct a Wilson
action~$S_\Lambda$ of QED with momentum cutoff~$\Lambda$ that keeps its BRST
invariance as we lower $\Lambda$. We derive the cutoff dependence of~$S_\Lambda$
as a differential equation, and also derive an expression for the BRST
invariance. The BRST transformation acts linearly on the action, and it is far
simpler than the BRST invariance of the Wilson action in the ERG formulation,
which is briefly reviewed in~Appendix~\ref{sec:C}. In~Sect.~\ref{sec:4} we
introduce a dimensionless framework by measuring dimensionful fields and
parameters in units of appropriate powers of the cutoff. We then construct the
BRST invariant Wilson action perturbatively in~Sect.~\ref{sec:5}. We only
consider the Wilson action for the continuum limit parametrized by the gauge
coupling, gauge fixing parameter, and the electron mass parameter. Since the
ghost fields are free, we can reduce the BRST invariance to the Ward--Takahashi
(WT) identity. This WT identity can be interpreted as manifest gauge invariance
even though the transformation of the gauge field is somewhat
modified.\footnote{We ask the reader to bear with the overblown title of the
paper.} We construct the Wilson action satisfying the WT identity to second
order in the gauge coupling. We conclude the paper in~Sect.~\ref{sec:6}.

We work in the $D$-dimensional Euclidean space, where $D=4-\epsilon$. We use
the shorthand notation for the momentum integrals:
\begin{equation}
   \int_p\equiv\int\frac{d^Dp}{(2\pi)^D}.
\label{eq:(1.1)}
\end{equation}
We also use the convention that the momentum cutoff decreases along the flow of
the renormalization group, and the beta functions and anomalous dimensions may
have the opposite signs to what the reader is familiar with.

\section{Preparation}
\label{sec:2}
\subsection{ERG}
\label{sec:2.1}
We would like to review the essence of the exact renormalization group (ERG for
short). In one formulation of ERG we construct the Wilson action of a theory in
terms of a field satisfying a simple diffusion equation. The flow of the Wilson
action is generated by the diffusion of the field. For gauge theories, we can
replace the simple diffusion equation by a covariant diffusion equation that is
consistent with BRST invariance. The replacement results in the gradient flow
exact renormalization group (GFERG for short). We will introduce the BRST
covariant diffusion for QED in the next subsection.

Let $\phi(x)$ be a real scalar field renormalized at momentum scale~$\mu$, and
let $S[\phi]$ be its action. We introduce a diffused field $\phi(t;x)$ as the
solution of a simple diffusion equation
\begin{equation}
   \partial_t\phi(t;x)=\partial^2\phi(t;x)
\label{eq:(2.1)}
\end{equation}
satisfying the initial condition
\begin{equation}
   \phi(0;x)=\phi(x).
\label{eq:(2.2)}
\end{equation}
We would like to construct a Wilson action equivalent to~$S[\phi]$ in terms of
the diffused field~$\phi(t;x)$ instead of~$\phi(x)$. Let $\Lambda$ be a
momentum scale smaller than~$\mu$ given by
\begin{equation}
   t=\frac{1}{\Lambda^2}-\frac{1}{\mu^2}>0
\label{eq:(2.3)}
\end{equation}
so that
\begin{equation}
   \partial_t=\frac{\Lambda^2}{2}\left(-\Lambda\partial_\Lambda\right).
\label{eq:(2.4)}
\end{equation}
We introduce the Wilson action $S_\Lambda[\phi]$ for momentum cutoff~$\Lambda$
by
\begin{equation}
   \exp\left(S_\Lambda[\phi]\right)
   \equiv
   \int[d\phi']\,
   \exp\left\{
   -\frac{\Lambda^2}{2}\int_p
   \left[\phi(p)-z_\Lambda\phi_\Lambda'(p)\right]
   \left[\phi(-p)-z_\Lambda\phi_\Lambda'(p)\right]
   +S[\phi']
   \right\},
\label{eq:(2.5)}
\end{equation}
where $\phi_\Lambda'(p)$ is the Fourier transform of the diffused field
\begin{equation}
   \phi'(t;x)=\int_p\,e^{ipx}\phi_\Lambda'(p)
\label{eq:(2.6)}
\end{equation}
satisfying the initial condition
\begin{equation}
    \phi'(0;x)=\phi'(x).
\label{eq:(2.7)}
\end{equation}
In momentum space it is trivial to solve the diffusion equation to obtain
\begin{equation}
   \phi_\Lambda'(p)
   =e^{-p^2\left(\frac{1}{\Lambda^2}-\frac{1}{\mu^2}\right)}\phi'(p).
\label{eq:(2.8)}
\end{equation}

By construction, $\phi(p)$ equals $z_\Lambda\phi_\Lambda'(p)$ with a squared
fluctuation of order $1/\Lambda^2$. It is not exactly the same
as~$z_\Lambda\phi_\Lambda'(p)$, but it corresponds to it. The choice
of~$z_\Lambda$ is not unique. For example, we can determine $z_\Lambda$ to
normalize the kinetic term of the Wilson action~$S_\Lambda[\phi]$.

The $\Lambda$-dependence of the Wilson action~$S_\Lambda [\phi]$ is given by the
ERG differential equation:
\begin{equation}
   -\Lambda\frac{\partial}{\partial\Lambda}
   e^{S_\Lambda[\phi]}
   =\int_p\,\left[
   \left(\frac{2p^2}{\Lambda^2}-\gamma_\Lambda\right)\phi(p)
   \frac{\delta}{\delta\phi(p)}
   +\frac{1}{\Lambda^2}
   \left(\frac{2p^2}{\Lambda^2}-\gamma_\Lambda+1\right)
   \frac{\delta^2}{\delta\phi(p)\delta\phi(-p)}
   \right]e^{S_\Lambda[\phi]},
\label{eq:(2.9)}
\end{equation}
where $\gamma_\Lambda$ is defined by
\begin{equation}
   \gamma_\Lambda\equiv-\Lambda\frac{d}{d\Lambda}\ln z_\Lambda.
\label{eq:(2.10)}
\end{equation}

For the correlation functions, we can give a precise relation between
$S_\Lambda$ and~$S$:
\begin{align}
   &\left\langle
   \exp\left[
   -\frac{1}{2\Lambda^2}\int_p\,
   \frac{\delta^2}{\delta\phi (p)\delta\phi(-p)}\right]
   \phi(p_1)\dotsb\phi(p_n)\right\rangle_{S_\Lambda}
\notag\\
   &=z_\Lambda^n
   \left\langle
   \phi_\Lambda(p_1)\dotsb\phi_\Lambda(p_n)\right\rangle_S
   =z_\Lambda^n
   \prod_{i=1}^n
   e^{-p_i^2\left(\frac{1}{\Lambda^2}-\frac{1}{\mu^2}\right)}
   \cdot\left\langle\phi(p_1)\dotsb\phi(p_n)\right\rangle_S.
\label{eq:(2.11)}
\end{align}
In constructing $S_\Lambda$, we have ``scrambled'' the field $\phi(p)$
around~$z_\Lambda\phi_\Lambda(p)$. We need to unscramble the field to get back
the same correlation functions. This is the role played by the exponentiated
differential operator.

An alternative definition of the Wilson action is given
by~\cite{Sonoda:2020vut}
\begin{equation}
   \exp\left(S_\Lambda[\phi]\right)
   =\Hat{s}_\Lambda
   \int[d\phi']\,
   \prod_p\delta\left(\phi(p)-z_\Lambda\phi_\Lambda'(p)\right)
   \cdot\exp\left(S[\phi']\right),
\label{eq:(2.12)}
\end{equation}
where
\begin{equation}
   \Hat{s}_\Lambda
   \equiv\exp\left[
   \frac{1}{2\Lambda^2}\int_p\,
   \frac{\delta^2}{\delta\phi(p)\delta\phi(-p)}\right]
\label{eq:(2.13)}
\end{equation}
is what we call the scrambler. The scrambler is necessary to maintain the
locality of the Wilson action~$S_\Lambda[\phi]$. The equality of the correlation
functions can be written as
\begin{equation}
   \left\langle
   \Hat{s}_\Lambda^{-1}
   \left[\phi(p_1)\dotsb\phi(p_n)\right]\right\rangle_{S_\Lambda}
   =z_\Lambda^n
   \left\langle\phi_\Lambda(p_1)\dotsb\phi_\Lambda(p_n)\right\rangle_S
\label{eq:(2.14)}
\end{equation}
using the unscrambler, i.e., the inverse of the scrambler.

\subsection{Diffusion in QED}
\label{sec:2.2}
In constructing the Wilson action of a gauge theory, we can use the simple
diffusion equation for gauge fields and matter as explained above for the
scalar theory. The Wilson action, thus constructed, retains gauge invariance
(BRST invariance to be more precise), but its realization is not as
straightforward as we wish~\cite{Igarashi:2009tj}. This is partially due to the
use of a simple diffusion of fields which does not respect the gauge
invariance. In~Ref.~\cite{Sonoda:2020vut} we have introduced an alternative
Wilson action based upon a covariant diffusion of fields, consistent with the
gauge invariance of the theory.\footnote{Preceding Ref.~\cite{Sonoda:2020vut},
other gauge invariant ERG formulations had been introduced
in~Refs.~\cite{Morris:1998kz,Morris:1999px,Morris:2000fs,Arnone:2005fb,%
Morris:2006in,Wetterich:2016ewc,Wetterich:2017aoy}.} In this subsection we
would like to introduce such diffusion explicitly for the simple case of QED.

Let us consider QED renormalized at momentum scale~$\mu$ in~$D=4-\epsilon$
dimensional Euclidean space. We denote the gauge field by~$A_\mu(x)$, and the
electron field by~$\psi(x)$ and~$\Bar{\psi}(x)$, and the free Faddeev--Popov
ghost fields by~$c(x)$ and~$\Bar{c}(x)$. The dimensionless gauge coupling
renormalized at~$\mu$ is~$e$. The action~$S$ is invariant under the following
BRST transformation of the renormalized fields:
\begin{subequations}\label{eq:(2.15)}
\begin{align}
   \delta A_\mu(x)
   &=\eta\partial_\mu c(x),
\\
   \delta c(x)&=0,
\\
   \delta\Bar{c}(x)
   &=\eta\frac{1}{\xi}\partial\cdot A(x),
\\
   \delta\psi(x)
   &=ie\mu^{\epsilon/2}\eta c(x)\psi(x),
\\
   \delta\Bar{\psi}(x)
   &=-ie\mu^{\epsilon/2}\eta c(x)\Bar{\psi}(x),
\end{align}
\end{subequations}
where $\eta$ is an arbitrary anticommuting number that keeps the statistics of
the fields under the transformation.\footnote{We have chosen the mass dimension
of~$c(x)$ as~$(D-4)/2=-\epsilon/2$ and that of~$\Bar{c}(x)$ as~$D/2$ so that
the mass dimension of~$\eta$ is zero.}

We introduce the following diffusion equations:
\begin{subequations}\label{eq:(2.16)}
\begin{align}
   \partial_t A_\mu(t;x)
   &=\partial^2 A_\mu(t;x),
\\
   \partial_t c(t;x)
   &=\partial^2c(t;x),
\\
   \partial_t\Bar{c}(t;x)
   &=\partial^2\Bar{c}(t;x),
\\
   \partial_t\psi(t;x)
   &=\left[
   \partial^2-2ie\mu^{\epsilon/2}A_\mu(t;x)\partial_\mu
   -e^2\mu^\epsilon A_\mu(t;x)A_\mu(t;x)
   \right]\psi(t;x),
\\
   \partial_t\Bar{\psi}(t;x)
   &=\left[
   \partial^2+2ie\mu^{\epsilon/2}A_\mu(t;x)\partial_\mu
   -e^2\mu^\epsilon A_\mu(t;x)A_\mu(t;x)
   \right]\Bar{\psi}(t;x),
\end{align}
\end{subequations}
where the fields match the renormalized fields at~$t=0$:
\begin{equation}
   A_\mu(0;x)=A_\mu(x),\dotsc,\Bar{\psi}(0;x)=\Bar{\psi}(x).
\label{eq:(2.17)}
\end{equation}

It is straightforward to check that the above diffusion equations are
consistent with the BRST transformation. Namely, the BRST transformation of the
fields at~$t=0$ implies the same BRST transformation of the diffused fields:
\begin{subequations}\label{eq:(2.18)}
\begin{align}
   \delta A_\mu(t;x)
   &=\eta\partial_\mu c(t;x),
\\
   \delta c(t;x)&=0,
\\
   \delta\Bar{c}(t;x)
   &=\eta\frac{1}{\xi}\partial\cdot A(t;x),
\\
   \delta\psi(t;x)
   &=ie\mu^{\epsilon/2}\eta c(t;x)\psi(t;x),
\\
   \delta\Bar{\psi}(t;x)
   &=-ie\mu^{\epsilon/2}\eta c(t;x)\Bar{\psi}(t;x).
\end{align}
\end{subequations}
To show this, we need to check that $\delta$ commutes with~$\partial_t$. Let us
check only two here.
\begin{equation}
   \delta\partial_t A_\mu(t;x)
   =\delta\partial^2A_\mu(t;x)
   =\partial^2\delta A_\mu(t;x)
   =\eta\partial^2\partial_\mu c(t;x)
\label{eq:(2.19)}
\end{equation}
is consistent with
\begin{equation}
   \partial_t\delta A_\mu(t;x)
   =\partial_t\eta\partial_\mu c(t;x)
   =\eta\partial_\mu\partial_t c(t;x)
   =\eta\partial_\mu\partial^2 c(t;x).
\label{eq:(2.20)}
\end{equation}
We also find
\begin{align}
   \delta\partial_t\psi(t;x)
   &=\delta\left[
   \partial^2-2ie\mu^{\epsilon/2}A_\mu(t;x)\partial_\mu
   -e^2\mu^\epsilon A_\mu(t;x)A_\mu(t;x)
   \right]\psi(t;x)
\notag\\
   &=\left[
   \partial^2-2ie\mu^{\epsilon/2}A_\mu(t;x)\partial_\mu
   -e^2\mu^\epsilon A_\mu(t;x)A_\mu(t;x)
   \right]
   \left[\eta ie\mu^{\epsilon/2}c(t;x)\psi(t;x)\right]
\notag\\
   &\qquad{}
   +\left[
   -2ie\mu^{\epsilon/2}\eta\partial_\mu c(t;x)\partial_\mu
   -2e^2\mu^\epsilon A_\mu(t;x)\eta\partial_\mu c(t;x)
   \right]\psi(t;x)
\notag\\
   &=\eta ie\mu^{\epsilon/2}\partial^2c(t;x)\cdot\psi(t;x)
\notag\\
   &\qquad{}
   +\eta ie\mu^{\epsilon/2}c(t;x)
   \left[
   \partial^2-2ie\mu^{\epsilon/2} A_\mu(t;x)\partial_\mu
   -e^2\mu^\epsilon A_\mu(t;x)A_\mu(t;x)
   \right]\psi(t;x)
\label{eq:(2.21)}
\end{align}
is consistent with
\begin{align}
   \partial_t\delta\psi(t;x)
   &=\partial_t\left[\eta ie\mu^{\epsilon/2}c(t;x)\psi(t;x)\right]
\notag\\
   &=\eta ie\mu^{\epsilon/2}\partial^2 c(t;x)\psi(t;x)
\notag\\
   &\qquad{}
   +\eta ie\mu^{\epsilon/2}c(t;x)
   \left[\partial^2-2ie\mu^{\epsilon/2}A_\mu(t;x)\partial_\mu
   -e^2\mu^\epsilon A_\mu(t;x)A_\mu(t;x)
   \right]\psi(t;x).
\label{eq:(2.22)}
\end{align}

We have thus introduced diffusion of fields preserving the form of the BRST
transformation. Our aim is to construct a Wilson action of QED using the
diffused fields as the elementary fields.

\section{GFERG for QED}
\label{sec:3}
We introduce the Wilson action of QED as
\begin{align}
   &e^{S_\Lambda[A_\mu,c,\Bar{c},\psi,\Bar{\psi}]}
\notag\\
   &\equiv\int\left[dA_\mu'dc'd\Bar{c}'d\psi'd\Bar{\psi}'\right]\,
   \notag\\
   &\qquad\qquad{}
   \times\exp
   \biggl[
   -\frac{\Lambda^2}{2}\int d^Dx\,
   \left(A_\mu-z_\Lambda A_{\Lambda \mu}'\right)^2
   -\Lambda^2\int d^D x\,
   \left(\Bar{c}-\Bar{c}_\Lambda' \right)\left(c-c_\Lambda'\right)
\notag\\
   &\qquad\qquad\qquad\qquad{}
   +i\Lambda\int d^Dx\,
   \left(\Bar{\psi}-z_{F\Lambda}\Bar{\psi}_\Lambda'\right)
   \left(\psi-z_{F\Lambda}\psi_\Lambda'\right)
   +S\left[A_\mu',c',\Bar{c}',\psi',\Bar{\psi}'\right]
   \biggr],
\label{eq:(3.1)}
\end{align}
where $z_\Lambda$ and~$z_{F\Lambda}$ satisfy
\begin{subequations}\label{eq:(3.2)}
\begin{align}
   -\Lambda\frac{\partial}{\partial\Lambda}\ln z_\Lambda
   &=\gamma_\Lambda,
\\
   -\Lambda\frac{\partial}{\partial\Lambda}\ln z_{F\Lambda}
   &=\gamma_{F\Lambda}.
\end{align}
\end{subequations}
We do not introduce any factor of wave function renormalization for the ghosts
since they remain free fields (see below). The diffused fields satisfy
\begin{subequations}\label{eq:(3.3)}
\begin{align}
   -\Lambda\partial_\Lambda A_{\Lambda\mu}
   &=\frac{2}{\Lambda^2}\partial^2A_{\Lambda\mu},
\\
   -\Lambda\partial_\Lambda c_\Lambda
   &=\frac{2}{\Lambda^2}\partial^2c_\Lambda,
\\
   -\Lambda\partial_\Lambda\Bar{c}_\Lambda
   &=\frac{2}{\Lambda^2}\partial^2\Bar{c}_\Lambda,
\\
   -\Lambda\partial_\Lambda\psi_\Lambda
   &=\frac{2}{\Lambda^2}
   \left(\partial^2-2ie\mu^{\epsilon/2}A_{\Lambda\mu}\partial_\mu
   -e^2\mu^\epsilon A_{\Lambda\mu}A_{\Lambda\mu}
   \right)\psi_\Lambda,
\\
   -\Lambda\partial_\Lambda\Bar{\psi}_\Lambda
   &=\frac{2}{\Lambda^2}
   \left(\partial^2+2ie\mu^{\epsilon/2}A_{\Lambda\mu}\partial_\mu
   -e^2\mu^\epsilon A_{\Lambda\mu}A_{\Lambda\mu}
   \right)\Bar{\psi}_\Lambda.
\end{align}
\end{subequations}

\subsection{GFERG differential equation}
\label{sec:3.1}
We wish to obtain
\begin{equation}
   -\Lambda\partial_\Lambda e^{S_\Lambda[A_\mu,c,\Bar{c},\psi,\Bar{\psi}]}
\label{eq:(3.4)}
\end{equation}
in terms of the functional derivatives with respect to the field variables. As
a preparation, we note the following correspondence:
\begin{subequations}\label{eq:(3.5)}
\begin{align}
   -\frac{1}{\Lambda^2}\frac{\delta}{\delta A_\mu(x)}
   &\longleftrightarrow A_\mu(x)-z_\Lambda A_{\Lambda\mu}'(x),
\label{eq:(3.5a)}\\
   -\frac{1}{\Lambda^2}\frac{\overrightarrow{\delta}}{\delta\Bar{c}(x)}
   &\longleftrightarrow c(x)-c_\Lambda'(x),
\\
   -\frac{1}{\Lambda^2}\frac{\overleftarrow{\delta}}{\delta c(x)}
   &\longleftrightarrow\Bar{c}(x)-\Bar{c}_\Lambda'(x),
\\
   -\frac{i}{\Lambda}\frac{\overleftarrow{\delta}}{\delta\psi(x)}
   &\longleftrightarrow\Bar{\psi}(x)-z_{F\Lambda}\Bar{\psi}_\Lambda'(x),
\\
   -\frac{i}{\Lambda}\frac{\overrightarrow{\delta}}{\delta\Bar{\psi}(x)}
   &\longleftrightarrow\psi(x)-z_{F\Lambda}\psi_\Lambda'(x).
\end{align}
\end{subequations}
We need to be clear about what we mean by the above correspondence. Take
Eq.~\eqref{eq:(3.5a)}. The correspondence means
\begin{align}
   &-\frac{1}{\Lambda^2}\frac{\delta}{\delta A_\mu(x)}e^{S_\Lambda}
\notag\\
   &=\int\left[dA_\mu'dc'd\Bar{c}'d\psi'd\Bar{\psi}'\right]\,
   \left[A_\mu(x)-z_\Lambda A_{\Lambda\mu}'(x)\right]
   \exp\left(
   \text{quadratic terms}+S\left[A_\mu',c',\Bar{c}',\psi',\Bar{\psi}'\right]
   \right).
\label{eq:(3.6)}
\end{align}
Differentiating this once more we obtain
\begin{align}
   &(-)\frac{1}{\Lambda^2}\frac{\delta}{\delta A_\mu(x_2)}
   (-)\frac{1}{\Lambda^2}\frac{\delta}{\delta A_\mu(x_1)}e^{S_\Lambda}
\notag\\
   &=\int\left[dA_\mu'dc'd\Bar{c}'d\psi'd\Bar{\psi}'\right]\,
   \left[A_\mu(x_1)-z_\Lambda A_{\Lambda\mu}'(x_1)\right]
   \left[A_\mu(x_2)-z_\Lambda A_{\Lambda\mu}'(x_2)\right]
\notag\\
   &\qquad\qquad{}
   \times\exp\left(
   \text{quadratic terms}+S\left[A_\mu',c',\Bar{c}',\psi',\Bar{\psi}'\right]
   \right),
\label{eq:(3.7)}
\end{align}
where we have chosen $x_2\neq x_1$ so that the second differentiation with
respect to~$A_\mu(x_2)$ acts only on the exponential, but not on~$A_\mu(x_1)$ in
the integrand. Taking the limit $x_2\to x_1$, we obtain
\begin{align}
   &\lim_{x_2\to x_1}
   (-)\frac{1}{\Lambda^2}\frac{\delta}{\delta A_\mu(x_2)}
   (-)\frac{1}{\Lambda^2}\frac{\delta}{\delta A_\mu(x_1)}e^{S_\Lambda}
\notag\\
   &=\int\left[dA_\mu'dc'd\Bar{c}'d\psi'd\Bar{\psi}'\right]\,
   \left[A_\mu(x_1)-z_\Lambda A_{\Lambda\mu}'(x_1)\right]
   \left[A_\mu(x_1)-z_\Lambda A_{\Lambda\mu}'(x_1)\right]
\notag\\
   &\qquad\qquad{}
   \times\exp\left(
   \text{quadratic terms}+S\left[A_\mu',c',\Bar{c}',\psi',\Bar{\psi}'\right]
   \right).
\label{eq:(3.8)}
\end{align}
Similarly, we obtain
\begin{align}
   &\lim_{x'\to x}
   (-)\frac{1}{\Lambda^2}\frac{\overrightarrow{\delta}}{\delta\Bar{c}(x)}
   e^{S_\Lambda}(-)\frac{1}{\Lambda^2}\frac{\overleftarrow{\delta}}{\delta c(x')}
\notag\\
   &=\int\left[dA_\mu'dc'd\Bar{c}'d\psi'd\Bar{\psi}'\right]\,
   \left[c(x)-c_\Lambda'(x)\right]
   \left[\Bar{c}(x)-\Bar{c}'(x)\right]
\notag\\
   &\qquad\qquad{}
   \times\exp\left(
   \text{quadratic terms}+S\left[A_\mu',c',\Bar{c}',\psi',\Bar{\psi}'\right]
   \right),
\label{eq:(3.9)}
\end{align}
where we take the limit $x'\to x$ after the differentiation.

We now calculate
\begin{align}
   &-\Lambda\frac{\partial}{\partial\Lambda}e^{S_\Lambda}
\notag\\
   &=\int\left[dA_\mu'dc'd\Bar{c}'d\psi'd\Bar{\psi}'\right]\,
   e^{\text{quadratic terms}+S\left[A_\mu',c',\Bar{c}',\psi',\Bar{\psi}'\right]}
\notag\\
   &\qquad{}
   \times\int d^Dx\,
   \biggl\{
   \Lambda^2\left(A_\mu-z_\Lambda A_{\Lambda\mu}'\right)^2
   +2\Lambda^2\left(\Bar{c}-\Bar{c}_\Lambda'\right)\left(c-c_\Lambda'\right)
   -i\Lambda\left(\Bar{\psi}-z_{F\Lambda}\Bar{\psi}_\Lambda'\right)
   \left(\psi-z_{F\Lambda}\psi_\Lambda'\right)
\notag\\
   &\qquad\qquad\qquad\qquad{}
   +\Lambda^2\left(A_\mu-z_\Lambda A_{\Lambda\mu}'\right)
   z_\Lambda\left(
   \gamma_\Lambda A_{\Lambda\mu}'+\frac{2}{\Lambda^2}\partial^2A_{\Lambda\mu}'
   \right)
\notag\\
   &\qquad\qquad\qquad\qquad{}
   +2\partial^2\Bar{c}_\Lambda'\left(c-c_\Lambda'\right)
   +2\left(\Bar{c}-\Bar{c}_\Lambda'\right)\partial^2c_\Lambda'
\notag\\
   &\qquad\qquad\qquad\qquad{}
   -i\Lambda z_{F\Lambda}
   \left[
   \gamma_{F\Lambda}
   +\frac{2}{\Lambda^2}
   \left(\partial^2+2ie\mu^{\epsilon/2}A_{\Lambda\mu}'\partial_\mu
   -e^2\mu^\epsilon A_{\Lambda\mu}'A_{\Lambda\mu}'
   \right)\right]
   \Bar{\psi}'
\notag\\
   &\qquad\qquad\qquad\qquad\qquad\qquad{}
   \times\left(\psi-z_{F\Lambda}\psi'\right)
\notag\\
   &\qquad\qquad\qquad\qquad{}
   -i\Lambda z_{F\Lambda}
   \left(\Bar{\psi}-z_{F\Lambda}\Bar{\psi}'\right)
\notag\\
   &\qquad\qquad\qquad\qquad\qquad\qquad{}
   \times\left[
   \gamma_{F\Lambda}
   +\frac{2}{\Lambda^2}
   \left(\partial^2-2ie\mu^{\epsilon/2}A_{\Lambda\mu}'\partial_\mu
   -e^2\mu^\epsilon A_{\Lambda\mu}'A_{\Lambda\mu}'
   \right)\right]
   \psi'\bigg\}.
\label{eq:(3.10)}
\end{align}
Using the correspondence given above, we can rewrite the rhs as
\begin{align}
   &-\Lambda\frac{\partial}{\partial\Lambda}e^{S_\Lambda}
\notag\\
   &=\int d^Dx\,
   \Biggl(
   \frac{1}{\Lambda^2}\frac{\delta^2}{\delta A_\mu(x)\delta A_\mu(x')}
   e^{S_\Lambda}
   -\frac{2}{\Lambda^2}\frac{\overrightarrow{\delta}}{\delta\Bar{c}(x)}
   e^{S_\Lambda}
   \frac{\overleftarrow{\delta}}{\delta c(x')}
   -\frac{i}{\Lambda}
   \Tr\frac{\overrightarrow{\delta}}{\delta\Bar{\psi}(x)}
   e^{S_\Lambda}
   \frac{\overleftarrow{\delta}}{\delta\psi(x')}
\notag\\
   &\qquad\qquad\qquad{}
   -\frac{2}{\Lambda^2}
   \partial^2\left[
   A_\mu(x)+\frac{1}{\Lambda^2}\frac{\delta}{\delta A_\mu(x)}\right]
   \frac{\delta}{\delta A_\mu(x')}
   e^{S_\Lambda}
\notag\\
   &\qquad\qquad\qquad{}
   +\frac{2}{\Lambda^2}\frac{\overrightarrow{\delta}}{\delta\Bar{c}(x')}
   e^{S_\Lambda}
   \partial^2\left[
   \Bar{c}(x)+\frac{1}{\Lambda^2}\frac{\overleftarrow{\delta}}{\delta c(x)}
   \right]
   +\frac{2}{\Lambda^2}\partial^2
   \left[
   c(x)+\frac{1}{\Lambda^2}\frac{\overrightarrow{\delta}}{\delta\Bar{c}(x)}
   \right]
   e^{S_\Lambda}
   \frac{\overleftarrow{\delta}}{\delta c(x')}
\notag\\
   &\qquad\qquad\qquad{}
   +\Tr\frac{\overrightarrow{\delta}}{\delta\Bar{\psi}(x')}
   e^{S_\Lambda}
   \Biggl\{
   \frac{2}{\Lambda^2}\partial^2
   +\frac{4ie_\Lambda}{\Lambda^2}
   \left(
   A_\mu+\frac{1}{\Lambda^2}\frac{\overleftarrow{\delta}}{\delta A_\mu}\right)
   \partial_\mu
\notag\\
   &\qquad\qquad\qquad\qquad\qquad\qquad\qquad{}
   -\frac{2e_\Lambda^2}{\Lambda^2}
   \left(
   A_\mu+\frac{1}{\Lambda^2}\frac{\overleftarrow{\delta}}{\delta A_\mu}
   \right)
   \left[
   A_\mu(x')+\frac{1}{\Lambda^2}\frac{\overleftarrow{\delta}}{\delta A_\mu(x')}
   \right]
   \Biggr\}
\notag\\
   &\qquad\qquad\qquad\qquad\qquad\qquad\qquad{}
   \times
   \left[\Bar{\psi}(x)+\frac{i}{\Lambda}
   \frac{\overleftarrow{\delta}}{\delta\psi(x)}\right]
\notag\\
   &\qquad\qquad\qquad{}
   +\Tr\Biggl\{
   \frac{2}{\Lambda^2}\partial^2
   -\frac{4ie_\Lambda}{\Lambda^2}
   \left(A_\mu+\frac{1}{\Lambda^2}\frac{\delta}{\delta A_\mu}\right)
   \partial_\mu
\notag\\
   &\qquad\qquad\qquad\qquad\qquad{}
   -\frac{2e_\Lambda^2}{\Lambda^2}
   \left(A_\mu+\frac{1}{\Lambda^2}\frac{\delta}{\delta A_\mu}\right)
   \left[A_\mu(x')+\frac{1}{\Lambda^2}\frac{\delta}{\delta A_\mu(x')}\right]
   \Biggr\}
\notag\\
   &\qquad\qquad\qquad\qquad\qquad{}
   \times\left[
   \psi(x)
   +\frac{i}{\Lambda}\frac{\overrightarrow{\delta}}{\delta\Bar{\psi(x)}}\right]
   e^{S_\Lambda}\frac{\overleftarrow{\delta}}{\delta\psi(x')}
\notag\\
   &\qquad\qquad\qquad{}
   -\gamma_\Lambda
   \left(A_\mu+\frac{1}{\Lambda^2}\frac{\delta}{\delta A_\mu}\right)
   \frac{\delta}{\delta A_\mu}e^{S_\Lambda}
\notag\\
   &\qquad\qquad\qquad{}
   +\gamma_{F\Lambda}\Tr\frac{\overrightarrow{\delta}}{\delta\Bar{\psi}(x')}
   e^{S_\Lambda}
   \left[
   \Bar{\psi}(x)+\frac{i}{\Lambda}\frac{\overleftarrow{\delta}}{\delta\psi(x)}
   \right]
\notag\\
   &\qquad\qquad\qquad{}
   +\gamma_{F\Lambda}\Tr\left[
   \psi(x)+\frac{i}{\Lambda}\frac{\overrightarrow{\delta}}{\delta\Bar{\psi}(x)}
   \right]e^{S_\Lambda}\frac{\overleftarrow{\delta}}{\delta\psi(x')}\Biggr)
\notag\\
   &=\int d^Dx\,
   \Biggl\{
   \biggl[
   \left(-\frac{2}{\Lambda^2}\partial^2-\gamma_\Lambda\right)
   A_\mu(x)\cdot\frac{\delta}{\delta A_\mu(x)}
\notag\\
   &\qquad\qquad\qquad{}
   +\left(-\frac{2}{\Lambda^2}\partial^2-\gamma_\Lambda+1\right)
   \frac{1}{\Lambda^2}\frac{\delta^2}{\delta A_\mu(x)\delta A_\mu(x')}
   \biggr]e^{S_\Lambda}
\notag\\
   &\qquad\qquad\qquad{}
   +\frac{2}{\Lambda^2}
   \left[-\partial^2\Bar{c}(x)\frac{\overrightarrow{\delta}}{\delta\Bar{c}(x)}
   e^{S_\Lambda}
   -e^{S_\Lambda}\frac{\overleftarrow{\delta}}{\delta c(x)}\partial^2c(x)\right]
\notag\\
   &\qquad\qquad\qquad{}
   +\left[
   \frac{1}{\Lambda^2}\left(\partial^2+\partial^{\prime2}\right)-1\right]
   \frac{2}{\Lambda^2}\frac{\overrightarrow{\delta}}{\delta\Bar{c}(x)}
   e^{S_\Lambda}\frac{\overleftarrow{\delta}}{\delta c(x')}
\notag\\
   &\qquad\qquad\qquad{}
   +\Tr\frac{\overrightarrow{\delta}}{\delta\Bar{\psi}(x')}
   e^{S_\Lambda}
   \left(\frac{2}{\Lambda^2}\partial^2+\gamma_{F\Lambda}\right)
   \left[
   \Bar{\psi}(x)+\frac{i}{\Lambda}\frac{\overleftarrow{\delta}}{\delta\psi(x)}
   \right]
\notag\\
   &\qquad\qquad\qquad{}
   +\Tr\left(\frac{2}{\Lambda^2}\partial^2+\gamma_{F\Lambda}\right)
   \left[
   \psi(x)+\frac{i}{\Lambda}\frac{\overrightarrow{\delta}}{\delta\Bar{\psi}(x)}
   \right]
   e^{S_\Lambda}
   \frac{\overleftarrow{\delta}}{\delta\psi(x')}
\notag\\
   &\qquad\qquad\qquad{}
   -\frac{i}{\Lambda}\Tr\frac{\overrightarrow{\delta}}{\delta\Bar{\psi}(x)}
   e^{S_\Lambda}\frac{\overleftarrow{\delta}}{\delta\psi(x')}
   \Biggr\}
\notag\\
   &\qquad{}
   +\int d^Dx\,
   \Tr\Biggl(
   \frac{\overrightarrow{\delta}}{\delta\Bar{\psi}(x')}
   e^{S_\Lambda}
   \Biggl\{\frac{4ie_\Lambda}{\Lambda^2}
   \left(
   A_\mu+\frac{1}{\Lambda^2}\frac{\overleftarrow{\delta}}{\delta A_\mu}\right)
   \partial_\mu
\notag\\
   &\qquad\qquad\qquad\qquad{}
   -\frac{2e_\Lambda^2}{\Lambda^2}
   \left(
   A_\mu+\frac{1}{\Lambda^2}\frac{\overleftarrow{\delta}}{\delta A_\mu}\right)
   \left[
   A_\mu(x')+\frac{1}{\Lambda^2}\frac{\overleftarrow{\delta}}{\delta A_\mu(x')}
   \right]
   \Biggr\}
   \left[
   \Bar{\psi}(x)+\frac{i}{\Lambda}\frac{\overleftarrow{\delta}}{\delta\psi(x)}
   \right]
\notag\\
   &\qquad\qquad\qquad\qquad{}
   +\biggl\{
   -\frac{4ie_\Lambda}{\Lambda^2}
   \left(A_\mu+\frac{1}{\Lambda^2}\frac{\delta}{\delta A_\mu}\right)\partial_\mu
\notag\\
   &\qquad\qquad\qquad\qquad\qquad{}
   -\frac{2e_\Lambda^2}{\Lambda^2}
   \left(A_\mu+\frac{1}{\Lambda^2}\frac{\delta}{\delta A_\mu}\right)
   \left[A_\mu(x')+\frac{1}{\Lambda^2}\frac{\delta}{\delta A_\mu(x')}\right]
   \biggr\}
\notag\\
   &\qquad\qquad\qquad\qquad\qquad\qquad{}
   \times
   \left[
   \psi(x)+\frac{i}{\Lambda}\frac{\overrightarrow{\delta}}{\delta\Bar{\psi(x)}}
   \right]
   e^{S_\Lambda}
   \frac{\overleftarrow{\delta}}{\delta\psi(x')}\Biggr),
\label{eq:(3.11)}
\end{align}
where the limit $x'\to x$ is implied, and we have defined the gauge coupling of
mass dimension~$\epsilon/2$ by
\begin{equation}
   e_\Lambda\equiv\frac{e\mu^{\epsilon/2}}{z_\Lambda}.
\label{eq:(3.12)}
\end{equation}
Note we have given Eq.~\eqref{eq:(3.11)} in two parts. The first part
reproduces the ERG differential equation. (See Appendix~\ref{sec:C} for a quick
review of ERG for QED.) The second part is unique to GFERG; it comes from the
BRST covariance of the electron diffusion equations~\eqref{eq:(3.3)}.

\subsection{BRST invariance}
\label{sec:3.2}
We next derive the expression of BRST invariance of $S_\Lambda$, inherited from
the invariance of~$S[A_\mu',c',\Bar{c}',\psi',\Bar{\psi}']$ under the BRST
transformation:
\begin{subequations}\label{eq:(3.13)}
\begin{align}
   \delta A'_\mu
   &=\eta\partial_\mu c',
\\
   \delta c'&=0,
\\
   \delta\Bar{c}'
   &=\eta\frac{1}{\xi}\partial_\mu A'_\mu,
\\
   \delta\psi'
   &=\eta ie\mu^{\epsilon/2}c'\psi',
\\
   \delta\Bar{\psi}'
   &=\eta(-i)e\mu^{\epsilon/2}c'\Bar{\psi}'.
\end{align}
\end{subequations}
As explained in~Sect.~\ref{sec:2.2}, this induces the diffused fields to
transform as
\begin{subequations}\label{eq:(3.14)}
\begin{align}
   \delta A_{\Lambda\mu}'
   &=\eta\partial_\mu c_\Lambda',
\\
   \delta c_\Lambda'&=0,
\\
   \delta\Bar{c}_\Lambda'
   &=\eta\frac{1}{\xi}\partial_\mu A_{\Lambda\mu}',
\\
   \delta\psi_\Lambda'
   &=\eta ie\mu^{\epsilon/2}c_\Lambda'\psi_\Lambda',
\\
   \delta\Bar{\psi}_\Lambda'
   &=\eta(-i)e\mu^{\epsilon/2}c_\Lambda'\Bar{\psi}_\Lambda'.
\end{align}
\end{subequations}
Hence, we obtain
\begin{align}
   0&=\int\left[dA_\mu'dc'd\Bar{c}'d\psi'd\Bar{\psi}'\right]\,
   e^{\text{quadratic terms}+S[A_\mu',c',\Bar{c}',\psi',\Bar{\psi}']}
\notag\\
   &\qquad{}
   \times\int d^Dx\,
   \biggl[
   \Lambda^2z_\Lambda\eta\partial_\mu c_\Lambda'\cdot
   \left(
   A_\mu-z_\Lambda A_{\Lambda\mu}'\right)
   +\Lambda^2\frac{1}{\xi}\eta\partial_\mu A_{\Lambda\mu}'\cdot
   \left(c-c_\Lambda'\right)
\notag\\
   &\qquad\qquad\qquad\qquad{}
   +i\Lambda z_{F\Lambda}\eta ie\mu^{\epsilon/2}c_\Lambda'\Bar{\psi}_\Lambda'
   \left(\psi-z_{F\Lambda}\psi_\Lambda'\right)
\notag\\
   &\qquad\qquad\qquad\qquad{}
   +i\Lambda\left(\Bar{\psi}-z_{F\Lambda}\Bar{\psi}_\Lambda'\right)
   \left(-\eta z_{F\Lambda}ie\mu^{\epsilon/2}c_\Lambda' \psi_\Lambda'\right)
   \biggr].
\label{eq:(3.15)}
\end{align}
Dividing this by~$z_\Lambda$, and defining
\begin{equation}
   \xi_\Lambda\equiv\xi z_\Lambda^2,
\label{eq:(3.16)}
\end{equation}
we can rewrite this as
\begin{align}
   0&=\int\left[dA_\mu'dc'd\Bar{c}'d\psi'd\Bar{\psi}'\right]\,
   e^{\text{quadratic terms}+S[A_\mu',c',\Bar{c}',\psi',\Bar{\psi}']}
\notag\\
   &\qquad{}
   \times\int d^Dx\,
   \bigg[
   \Lambda^2\partial_\mu c_\Lambda'\cdot
   \left(
   A_\mu-z_\Lambda A_{\Lambda\mu}'\right)
   +\Lambda^2\frac{1}{\xi_\Lambda}z_\Lambda
   \partial_\mu A_{\Lambda\mu}'\cdot\left(c-c_\Lambda'\right)
\notag\\
   &\qquad\qquad\qquad\qquad{}
   +i\Lambda z_{F\Lambda}ie_\Lambda c_\Lambda'\Bar{\psi}_\Lambda'
   \left(\psi-z_{F\Lambda}\psi_\Lambda' \right)
   +i\Lambda\left(\Bar{\psi}-z_{F\Lambda}\Bar{\psi}_\Lambda'\right)
   z_{F\Lambda}ie_\Lambda c_\Lambda'\psi_\Lambda'
   \biggr],
\label{eq:(3.17)}
\end{align}
where we recall Eq.~\eqref{eq:(3.12)} defining $e_\Lambda$. Using the
differentials, we can rewrite this further as
\begin{align}
   &\int d^D x\,
   \Biggl\{
   -\partial_\mu\left[
   c(x)+\frac{1}{\Lambda^2}\frac{\overrightarrow{\delta}}{\delta\Bar{c}(x)}
   \right]
   \frac{\delta}{\delta A_\mu(x')}e^{S_\Lambda}
   -\frac{1}{\xi_\Lambda}
   \partial_\mu\left[
   A_\mu(x)+\frac{1}{\Lambda^2}\frac{\delta}{\delta A_\mu}(x)\right]
   \frac{\overrightarrow{\delta}}{\delta\Bar{c}(x')}e^{S_\Lambda}
\notag\\
   &\qquad\qquad{}
   -ie_\Lambda
   \left[
   c(x)+\frac{1}{\Lambda^2}\frac{\overrightarrow{\delta}}{\delta\Bar{c}(x)}
   \right]
   \Tr\frac{\overrightarrow{\delta}}{\delta\Bar{\psi}(x')}
   e^{S_\Lambda}
   \left[
   \Bar{\psi}(x)+\frac{i}{\Lambda}\frac{\overleftarrow{\delta}}{\delta\psi(x)}
   \right]
\notag\\
   &\qquad\qquad{}
   +ie_\Lambda
   \left[
   c(x)+\frac{1}{\Lambda^2}\frac{\overrightarrow{\delta}}{\delta\Bar{c}(x)}
   \right]
   \Tr\left[
   \psi(x)+\frac{i}{\Lambda}\frac{\overrightarrow{\delta}}{\delta\Bar{\psi}(x)}
   \right]
   e^{S_\Lambda}\frac{\overleftarrow{\delta}}{\delta\psi(x')}
   \Biggr\}=0.
\label{eq:(3.18)}
\end{align}
The terms with second order differentials with respect to~$\psi$
and~$\Bar{\psi}$ cancel, and we obtain finally
\begin{align}
   &\int d^D x\,
   \Biggl\{
   -\partial_\mu
   \left[
   c(x)+\frac{1}{\Lambda^2}\frac{\overrightarrow{\delta}}{\delta\Bar{c}(x)}
   \right]
   \frac{\delta}{\delta A_\mu(x)}e^{S_\Lambda}
   -\frac{1}{\xi_\Lambda}\partial_\mu
   \left[A_\mu(x)+\frac{1}{\Lambda^2}\frac{\delta}{\delta A_\mu(x)}\right]
   \frac{\overrightarrow{\delta}}{\delta\Bar{c}(x)}e^{S_\Lambda}
\notag\\
   &\qquad\qquad{}
   +ie_\Lambda
   \left[
   c(x)+\frac{1}{\Lambda^2}\frac{\overrightarrow{\delta}}{\delta\Bar{c}(x)}
   \right]\Bar{\psi}(x)
   \frac{\overrightarrow{\delta}}{\delta\Bar{\psi}(x)}e^{S_\Lambda}
\notag\\
   &\qquad\qquad{}
   -ie_\Lambda
   \left[
   c(x)+\frac{1}{\Lambda^2}\frac{\overrightarrow{\delta}}{\delta\Bar{c}(x)}
   \right]e^{S_\Lambda}
   \frac{\overleftarrow{\delta}}{\delta\psi(x)}\psi(x)
   \Biggr\}=0.
\label{eq:(3.19)}
\end{align}
This has come pretty close to manifest BRST invariance. The simple expression
is due to the BRST covariant diffusion of the fields, Eq.~\eqref{eq:(3.3)}.
In~Appendix~\ref{sec:C} we give the standard ERG formulation of QED for
comparison.

\section{GFERG and BRST in the dimensionless framework}
\label{sec:4}
To gain more insights into the scaling properties of the Wilson action, we
adopt the dimensionless framework. Instead of the momentum cutoff~$\Lambda$, we
use the dimensionless logarithmic scale parameter~$\tau$ defined by
\begin{equation}
   \Lambda=\mu e^{-\tau}.
\label{eq:(4.1)}
\end{equation}
We write the Fourier transforms of the fields in terms of the dimensionless
fields (with tildes) as
\begin{subequations}\label{eq:(4.2)}
\begin{align}
   A_\mu(k)&=\Lambda^{-(D+2)/2}\Tilde{A}_\mu\left(k/\Lambda\right),
\\
   c(k)&=\Lambda^{-(D+4)/2}\Tilde{c}\left(k/\Lambda\right),
\\
   \Bar{c}(-k)&=\Lambda^{-D/2}\Tilde{\Bar{c}}\left(-k/\Lambda\right),
\\
   \psi(p)&=\Lambda^{-(D+1)/2}\Tilde{\psi}\left(p/\Lambda\right),
\\
   \Bar{\psi}(-p)&=\Lambda^{-(D+1)/2}\Tilde{\Bar{\psi}}\left(-p/\Lambda\right).
\end{align}
\end{subequations}
The dimensionless gauge coupling is defined by
\begin{equation}
   \Tilde{e}_\tau
   \equiv\frac{e_\Lambda}{\Lambda^{\epsilon/2}}
   =\frac{e}{z_\Lambda}\left(\frac{\mu}{\Lambda}\right)^{\epsilon/2}.
\label{eq:(4.3)}
\end{equation}
The gauge fixing parameter remains the same:
\begin{equation}
   \Tilde{\xi}_\tau=\xi_\Lambda=\xi z_\Lambda^2.
\label{eq:(4.4)}
\end{equation}
Denoting
\begin{equation}
   \Tilde{\gamma}_\tau
   =\gamma_\Lambda
   =-\Lambda\frac{\partial}{\partial\Lambda}\ln z_\Lambda,
\label{eq:(4.5)}
\end{equation}
we obtain
\begin{align}
   \frac{d}{d\tau}\Tilde{e}_\tau
   &=\left(\frac{\epsilon}{2}-\Tilde{\gamma}_\tau\right)\Tilde{e}_\tau,
\label{eq:(4.6)}\\
   \frac{d}{d\tau}\Tilde{\xi}_\tau
   &=2\Tilde{\gamma}_\tau\Tilde{\xi}_\tau.
\label{eq:(4.7)}
\end{align}

The Wilson action in the dimensionless framework is a functional of 
the dimensionless fields above:
\begin{equation}
   \Tilde{S}_\tau\left[
   \Tilde{A}_\mu,\Tilde{c},\Tilde{\Bar{c}},\Tilde{\psi},\Tilde{\Bar{\psi}}
   \right]
   \equiv S_\Lambda\left[A_\mu,c,\Bar{c},\psi,\Bar{\psi}\right].
\label{eq:(4.8)}
\end{equation}
In order to rewrite the GFERG differential equation~\eqref{eq:(3.11)} in the
dimensionless framework, we need to use
\begin{align}
   &\frac{\partial}{\partial\tau} e^{\Tilde{S}_\tau}
\notag\\
   &=-\Lambda\frac{\partial}{\partial\Lambda}e^{S_\Lambda}
\notag\\
   &\qquad{}
   +\int_k\,
   \left(\frac{D+2}{2}+k\cdot\partial_k\right)
   \Tilde{A}_\mu(k)\cdot\frac{\delta}{\delta\Tilde{A}_\mu(k)}e^{\Tilde{S}_\tau}
\notag\\
   &\qquad{}
   +\int_k\,e^{\Tilde{S}_\tau}\frac{\overleftarrow{\delta}}{\delta\Tilde{c}(k)}
   \left(\frac{D+4}{2}+k\cdot\partial_k\right)\Tilde{c}(k)
   +\int_k\,\left(\frac{D}{2}+k\cdot\partial_k\right)
   \Tilde{\Bar{c}}(-k)
   \cdot\frac{\overrightarrow{\delta}}{\delta\Tilde{\Bar{c}} (-k)}
   e^{\Tilde{S}_\tau}
\notag\\
   &\qquad{}
   +\int_p\,e^{\Tilde{S}_\tau}\frac{\overleftarrow{\delta}}{\delta\Tilde{\psi}(p)}
   \left(\frac{D+1}{2}+p\cdot\partial_p\right)\Tilde{\psi}(p)
   +\int_p\,\left(\frac{D+1}{2}+p\cdot\partial_p\right)
   \Tilde{\Bar{\psi}}(-p)\cdot
   \frac{\overrightarrow{\delta}}{\delta\Tilde{\Bar{\psi}}(-p)}
   e^{\Tilde{S}_\tau}.
\label{eq:(4.9)}
\end{align}
Since we work only in the dimensionless framework from now, we omit the tildes
altogether, and we obtain the GFERG equation as
\begin{align}
   &\partial_\tau e^{S_\tau}
\notag\\
   &=\int_k\,
   \biggl[
   \left(2k^2+\frac{D+2}{2}-\gamma_\tau+k\cdot\partial_k\right)A_\mu(k)
   \cdot\frac{\delta}{\delta A_\mu(k)}e^{S_\tau}
\notag\\
   &\qquad\qquad{}
   +\left(2k^2+1-\gamma_\tau\right)
   \frac{\delta^2}{\delta A_\mu(k)\delta A_\mu(-k)}e^{S_\tau}
   \biggr]
\notag\\
   &\qquad{}
   +\int_k\,
   \Biggl[
   \left(2k^2+\frac{D}{2}+k\cdot\partial_k\right)\Bar{c}(-k)
   \cdot\frac{\overrightarrow{\delta}}{\delta\Bar{c}(-k)}e^{S_\tau}
\notag\\
   &\qquad\qquad\qquad{}
   +e^{S_\tau}\frac{\overleftarrow{\delta}}{\delta c(k)}
   \left(2k^2+\frac{D+4}{2}+k\cdot\partial_k\right)c(k)
   -2(2k^2+1)\frac{\overrightarrow{\delta}}{\delta\Bar{c}(-k)}e^{S_\tau}
   \frac{\overleftarrow{\delta}}{\delta c(k)}
   \Biggr]
\notag\\
   &\qquad{}
   +\int_p\,
   \Biggl[
   e^{S_\tau}\frac{\overleftarrow{\delta}}{\delta\psi(p)}
   \left(2p^2+\frac{D+1}{2}-\gamma_{F\tau}+p\cdot\partial_p\right)\psi(p)
\notag\\
   &\qquad\qquad\qquad{}
   +\left(
   2p^2+\frac{D+1}{2}-\gamma_{F\tau}+p\cdot\partial_p\right)\Bar{\psi}(-p)
   \cdot\frac{\overrightarrow{\delta}}{\delta\Bar{\psi}(-p)}e^{S_\tau}
\notag\\
   &\qquad\qquad\qquad{}
   -i\left(4p^2+1-2\gamma_{F\tau}\right) 
   \Tr\frac{\overrightarrow{\delta}}{\delta\Bar{\psi}(-p)}e^{S_\tau}
   \frac{\overleftarrow{\delta}}{\delta\psi(p)}
   \Biggr]
\notag\\
   &\qquad{}
   +\int d^Dx\,\Tr\frac{\overrightarrow{\delta}}{\delta\Bar{\psi}(x')}
   e^{S_\tau}
   \Biggl\{
   4ie_\tau\left[
   A_\mu(x)+\frac{\overleftarrow{\delta}}{\delta A_\mu(x)}\right]
   \partial_\mu
\notag\\
   &\qquad\qquad\qquad\qquad\qquad{}
   -2e_\tau^2
   \left[A_\mu(x)+\frac{\overleftarrow{\delta}}{\delta A_\mu(x)}\right]
   \left[A_\mu(x')+\frac{\overleftarrow{\delta}}{\delta A_\mu(x')}\right]
   \Biggr\}
   \left[\Bar{\psi}(x)+i\frac{\overleftarrow{\delta}}{\delta\psi(x)}\right]
\notag\\
   &\qquad{}
   +\int d^Dx\,\Tr\biggl\{
   -4ie_\tau\left[A_\mu(x)+\frac{\delta}{\delta A_\mu(x)}\right]
   \partial_\mu
\notag\\
   &\qquad\qquad\qquad{}
   -2e_\tau^2
   \left[A_\mu(x)+\frac{\delta}{\delta A_\mu(x)}\right]
   \left[A_\mu(x')+\frac{\delta}{\delta A_\mu(x')}\right]
   \biggl\}
   \left[\psi(x)+i\frac{\overrightarrow{\delta}}{\delta\Bar{\psi}(x)}\right]
   e^{S_\tau}\frac{\overleftarrow{\delta}}{\delta\psi(x')}.
\label{eq:(4.10)}
\end{align}
Except for the last two integrals, the rhs coincides with the ERG equation. For
the last two integrals we have kept the coordinate space notation to take the
limit~$x'\to x$ carefully.\footnote{The dimensionless coordinate is given by
$\Tilde{x}_\mu\equiv\Lambda x_\mu$. We have omitted the tilde in the GFERG
equation.}

Similarly, we can rewrite the BRST invariance in the dimensionless framework.
In coordinate space it is given by
\begin{align}
   &\int d^Dx\,
   \Biggl\{
   -\partial_\mu\left[
   c(x)+\frac{\overrightarrow{\delta}}{\delta\Bar{c}(x)}\right]
   \frac{\delta}{\delta A_\mu(x)}e^{S_\tau}
   -\frac{1}{\xi_\tau}\partial_\mu
   \left[A_\mu(x)+\frac{\delta}{\delta A_\mu(x)}\right]
   \frac{\overrightarrow{\delta}}{\delta\Bar{c}(x)}e^{S_\tau}
\notag\\
   &\qquad\qquad{}
   +ie_\tau\left[
   c(x)+\frac{\overrightarrow{\delta}}{\delta\Bar{c}(x)}\right]\Bar{\psi}(x)
   \frac{\overrightarrow{\delta}}{\delta\Bar{\psi}(x)}e^{S_\tau}
   -ie_\tau\left[c(x)+\frac{\overrightarrow{\delta}}{\delta\Bar{c}(x)}\right]
   e^{S_\tau}\frac{\overleftarrow{\delta}}{\delta\psi(x)}\psi(x)
   \Biggr\}=0.
\label{eq:(4.11)}
\end{align}

\section{Perturbative solution}
\label{sec:5}
In constructing the Wilson action~$S_\tau$, we assume that it does not depend
on~$\tau$ explicitly: we assume that its $\tau$ dependence comes only through
the $\tau$ dependence of three dimensionless parameters, i.e., the gauge
coupling~$e_\tau$, gauge fixing parameter~$\xi_\tau$, and the electron mass
parameter~$m_\tau$. This assumption is not valid, however, for~$S_\tau$ that is
the dimensionless version of~$S_\Lambda$ given by~Eq.~\eqref{eq:(3.1)}. As long
as the renormalization scale~$\mu$ is finite, $S_\Lambda$ depends
on~$\Lambda/\mu=e^{-\tau}$ explicitly. To remove this, we must take the
``continuum limit'', i.e., we must take~$\mu\to+\infty$.

Both the differential equation~\eqref{eq:(4.10)} and the BRST
invariance~\eqref{eq:(4.11)} have been derived based on the integral
formula~\eqref{eq:(3.1)}, but neither has explicit dependence on~$\tau$. In
practice we can construct the continuum limit of~$S_\tau$ by solving
Eqs.~\eqref{eq:(4.10)} and~\eqref{eq:(4.11)} simultaneously under the above
assumption. The gauge coupling~$e_\tau$ and the gauge fixing
parameter~$\xi_\tau$ are introduced through the BRST
invariance~\eqref{eq:(4.11)}. We normalize the kinetic terms of the gauge and
electron fields; this fixes the anomalous dimensions $\gamma_\tau$
and~$\gamma_{F\tau}$ as functions of~$e_\tau^2$ although they may also depend
on~$\xi_\tau$. (We believe neither depends on~$\xi_\tau$.)  We introduce a
normalization condition of the electron mass term that determines the $\tau$
dependence of~$m_\tau$ in the form
\begin{equation}
   \frac{d}{d\tau}m_\tau=\left[1+\beta_m(e_\tau^2)\right]m_\tau.
\label{eq:(5.1)}
\end{equation}
We believe $\beta_m$ is also independent of~$\xi_\tau$. We thus obtain
\begin{align}
   \partial_\tau e^{S_\tau}
   &=\partial_\tau e^{S(e_\tau,\xi_\tau,m_\tau)}
\notag\\
   &=\left\{
   \left[\frac{\epsilon}{2}-\gamma(e_\tau^2)\right]e_\tau
   \frac{\partial}{\partial e_\tau}
   +2\gamma(e_\tau^2)\xi_\tau\frac{\partial}{\partial\xi_\tau}
   +\left[1+\beta_m(e_\tau^2)\right]m_\tau
   \frac{\partial}{\partial m_\tau}
   \right\}
   e^{S(e_\tau,\xi_\tau,m_\tau)}.
\label{eq:(5.2)}
\end{align}
We will drop the suffix~$\tau$ from the parameters $e_\tau$, $\xi_\tau$,
and~$m_\tau$.\footnote{We write $e$ for~$e_\tau$. It should not be confused with
the renormalized coupling~$e$ of the original action~$S$.} Thus, our GFERG
differential equation becomes
\begin{align}
   &\left\{
   \left[\frac{\epsilon}{2}-\gamma(e^2)\right]e\frac{\partial}{\partial e}
   +2\gamma(e^2)\xi\frac{\partial}{\partial\xi}
   +\left[1+\beta_m(e^2)\right]m\frac{\partial}{\partial m}\right\}
   e^{S(e,\xi,m)}
\notag\\
   &=\text{rhs of Eq.~\eqref{eq:(4.10)}}.
\label{eq:(5.3)}
\end{align}
If we define the beta function of~$e^2$ by
\begin{equation}
   \beta(e^2)\equiv-2\gamma(e^2)e^2,
\label{eq:(5.4)}
\end{equation}
we can rewrite
\begin{equation}
   \left[\frac{\epsilon}{2}-\gamma(e^2)\right]e\frac{\partial}{\partial e}
   =\left[\frac{\epsilon}{2}+\frac{\beta(e^2)}{2e^2}\right]
   e\frac{\partial}{\partial e}.
\label{eq:(5.5)}
\end{equation}
Our purpose is to solve the GFERG equation~\eqref{eq:(5.3)} and the BRST
invariance~\eqref{eq:(4.11)} together perturbatively in powers of~$e$. For this
purpose we expand the Wilson action in powers of~$e$:
\begin{equation}
   S(e,\xi,m)=S^{(0)}(\xi,m)+eS^{(1)}(\xi,m)+e^2S^{(2)}(\xi, m)+\dotsb.
\label{eq:(5.6)}
\end{equation}

\subsection{Tree level}
\label{sec:5.1}
$S^{(0)}$ satisfies the GFERG equation
\begin{align}
   &m\partial_m S^{(0)}
\notag\\
   &=\int_k\,\left[
   \left(2k^2+\frac{D+2}{2}+k\cdot\partial_k\right)A_\mu(k)
   \cdot\frac{\delta S^{(0)}}{\delta A_\mu(k)}
   +2(2k^2+1)\frac{1}{2}
   \frac{\delta S^{(0)}}{\delta A_\mu(k)}\frac{\delta S^{(0)}}{\delta A_\mu(-k)}
   \right]
\notag\\
   &\qquad{}
   +\int_k\,
   \Biggl[
   S^{(0)}\frac{\overleftarrow{\delta}}{\delta c(k)}
   \left(2k^2+\frac{D+4}{2}+k\cdot\partial_k\right)c(k)
\notag\\
   &\qquad\qquad\qquad{}
   +\left(2k^2+\frac{D}{2}+k\cdot\partial_k\right)\Bar{c}(-k)
   \cdot\frac{\overrightarrow{\delta}}{\delta\Bar{c}(-k)}
    S^{(0)}
\notag\\
   &\qquad\qquad\qquad{}
   +2(2k^2+1)S^{(0)}\frac{\overleftarrow{\delta}}{\delta c(k)}
   \cdot\frac{\overrightarrow{\delta}}{\delta\Bar{c}(-k)}S^{(0)}
   \Biggr]
\notag\\ 
   &\qquad{}
   +\int_p\,
   \Biggl[S^{(0)}\frac{\overleftarrow{\delta}}{\delta\psi(p)}
   \left(2p^2+\frac{D+1}{2}+p\cdot\partial_p\right)\psi(p)
\notag\\
   &\qquad\qquad\qquad{}
   +\left(2p^2+\frac{D+1}{2}+p\cdot\partial_p\right)\Bar{\psi}(-p)
   \cdot\frac{\overrightarrow{\delta}}{\delta\Bar{\psi}(-p)}S^{(0)}
\notag\\
   &\qquad\qquad\qquad{}
   +i(4p^2+1)S^{(0)}\frac{\overleftarrow{\delta}}{\delta\psi(p)}
   \frac{\overrightarrow{\delta}}{\delta\Bar{\psi}(-p)}S^{(0)}
   \Biggr].
\label{eq:(5.7)}
\end{align}
This is solved by
\begin{align}
   S^{(0)}
   &=-\frac{1}{2}\int_k\,
   A_\mu(k)A_\nu(-k)
   \left[
   \left(\delta_{\mu\nu}-\frac{k_\mu k_\nu}{k^2}\right)\frac{k^2}{e^{-2k^2}+k^2}
   +\frac{k_\mu k_\nu}{k^2}\frac{k^2}{\xi e^{-2k^2}+k^2}
   \right]
\notag\\
   &\qquad{}
   -\int_k\,\Bar{c}(-k)c(k)\frac{k^2}{e^{-2k^2}+k^2}
   -\int_p\,\Bar{\psi}(-p)
   \frac{\Slash{p}+im}{e^{-2p^2}+i\left(\Slash{p}+im\right)}\psi(p).
\label{eq:(5.8)}
\end{align}
We have normalized the kinetic terms appropriately.

The $\xi$-dependence of the longitudinal part is determined by the BRST
invariance~\eqref{eq:(4.11)}; at tree level it gives
\begin{equation}
   \int_k\,\left\{
   k_\mu\left[
   c(k)+\frac{\overrightarrow{\delta}}{\delta\Bar{c}(-k)}S^{(0)}
   \right]
   \frac{\delta S^{(0)}}{\delta A_\mu(k)}
   -\frac{1}{\xi}k_\mu\left[
   A_\mu(-k)+\frac{\delta S^{(0)}}{\delta A_\mu(k)}\right]
   \frac{\overrightarrow{\delta}}{\delta\Bar{c}(-k)}S^{(0)}
   \right\}=0.
\label{eq:(5.9)}
\end{equation}
Let us check that Eq.~\eqref{eq:(5.8)} satisfies this:
\begin{subequations}\label{eq:(5.10)}
\begin{align}
   &\int_k\,
   k_\mu\left[
   c(k)+\frac{\overrightarrow{\delta}}{\delta\Bar{c}(-k)}S^{(0)}\right]
   \frac{\delta S^{(0)}}{\delta A_\mu(k)}
\notag\\
   &=\int_k\,c(k)
   \left(1-\frac{k^2}{e^{-2k^2}+k^2}\right)
   k_\mu A_\mu(-k)\frac{-k^2}{\xi e^{-2k^2}+k^2}
\notag\\
   &=\int_k\,c(k)
   \frac{e^{-2k^2}}{e^{-2k^2}+k^2}k_\mu A_\mu(-k)\frac{-k^2}{\xi e^{-2k^2}+k^2},
\\
   &\int_k\,\frac{1}{\xi}k_\mu
   \left[A_\mu(-k)+\frac{\delta S^{(0)}}{\delta A_\mu(k)}\right]
   \frac{\overrightarrow{\delta}}{\delta\Bar{c}(-k)}S^{(0)}
\notag\\
   &=\int_k\,\frac{1}{\xi}k_\mu A_\mu(-k)
   \left(1-\frac{k^2}{\xi e^{-2k^2}+k^2}\right)
   \frac{-k^2}{e^{-2k^2}+k^2}c(k)
\notag\\
   &=\int_k\,k_\mu A_\mu(-k)
   \frac{e^{-2k^2}}{\xi e^{-2k^2}+k^2}\frac{-k^2}{e^{-2k^2}+k^2}c(k).
\end{align}
\end{subequations}
Hence, $S^{(0)}$ is BRST invariant.

\subsection{BRST invariance simplified to manifest gauge invariance}
\label{sec:5.2}
Before proceeding to calculate $S^{(1)}$, we stop to simplify our expression for
the BRST invariance given in~Eq.~\eqref{eq:(4.11)}.

GFERG equation~\eqref{eq:(5.3)} implies the absence of higher order
corrections to the ghost dependent part of the action~$S$.\footnote{Recall
Eq.~\eqref{eq:(3.1)}. Since the original action $S$ is quadratic in ghost
fields, $S_\Lambda$ produced by the Gaussian functional integration of~$S$
remains quadratic in ghost fields.} Hence, the ghost part of the action is
exactly as given in~Eq.~\eqref{eq:(5.8)}:
\begin{equation}
   S_{\text{ghost}}=-\int_k\,\Bar{c}(-k)\frac{k^2}{e^{-2k^2}+k^2}c(k).
\label{eq:(5.11)}
\end{equation}
Hence, we can rewrite the BRST invariance as
\begin{align}
   &\int_k\,\Biggl\{
   (-ik_\mu)\frac{e^{-2k^2}}{e^{-2k^2}+k^2}c(k)\frac{\delta S}{\delta A_\mu(k)}
   +\frac{1}{\xi}ik_\mu
   \left[A_\mu(-k)+\frac{\delta S}{\delta A_\mu(k)}\right]
   \frac{-k^2}{e^{-2k^2}+k^2}c(k)
\notag\\
   &\qquad{}
   +ie\frac{e^{-2k^2}}{e^{-2k^2}+k^2}c(k)
   \int_p\,
   \Bar{\psi}(-p-k)\frac{\overrightarrow{\delta}}{\delta\Bar{\psi}(-p)}S
   -ie\frac{e^{-2k^2}}{e^{-2k^2}+k^2}c(k)
   \int_p\,S\frac{\overleftarrow{\delta}}{\delta\psi(p+k)}\psi(p)
   \Biggr\}
\notag\\
   &=0.
\label{eq:(5.12)}
\end{align}
The integrand is proportional to~$c(k)$, and its coefficient must vanish. This
results in the Ward-Takahashi (WT) identity given by
\begin{equation}
   \frac{\xi e^{-2k^2}+k^2}{\xi e^{-2k^2}}k_\mu
   \frac{\delta S_I}{\delta A_\mu(k)} 
   =e\int_p\,
   \left[
   \Bar{\psi}(-p-k)\frac{\overrightarrow{\delta}}{\delta\Bar{\psi}(-p)}S
   -S\frac{\overleftarrow{\delta}}{\delta\psi(p+k)}\psi(p)\right],
\label{eq:(5.13)}
\end{equation}
where
\begin{equation}
   S_I\equiv S-S^{(0)}
\label{eq:(5.14)}
\end{equation}
is the interaction part, and we have used the BRST invariance of~$S^{(0)}$.
Equation~\eqref{eq:(5.13)} differs from the classical gauge invariance
\begin{equation}
   k_\mu\frac{\delta S_{\text{classical}}}{\delta A_\mu(k)}
   =e\int_p\,
   \left[\Bar{\psi}(-p-k)\frac{\overrightarrow{\delta}}{\delta\Bar{\psi}(-p)}
   S_{\text{classical}}
   -S_{\text{classical}}\frac{\overleftarrow{\delta}}{\delta\psi(p+k)}\psi(p)
   \right]
\label{eq:(5.15)}
\end{equation}
merely by the $k$-dependent factor on the lhs.

In fact Eq.~\eqref{eq:(5.13)} implies that the action
\begin{equation}
   S_{\text{inv}}\equiv S
   +\frac{1}{2}\int_k\,A_\mu(k)A_\nu(-k)\frac{k_\mu k_\nu}{\xi e^{-2k^2}+k^2}
   +\int_k\,\bar{c}(-k)\frac{k^2}{e^{-2k^2}+k^2}c(k)
\label{eq:(5.16)}
\end{equation}
without the gauge fixing and ghost terms is invariant under the infinitesimal
``gauge'' transformation
\begin{subequations}\label{eq:(5.17)}
\begin{align}
   \delta A_\mu(k)
   &=-\frac{\xi e^{-2k^2}+k^2}{\xi e^{-2k^2}}k_\mu\chi(k),
\\
   \delta\psi(p)
   &=-e\int_k\,\chi(k)\psi(p-k),
\\
   \delta\Bar{\psi}(-p)
   &=e\int_k\,\chi(k)\Bar{\psi}(-p-k),
\end{align}
\end{subequations}
where $\chi(k)$ is an arbitrary infinitesimal function. On this account we may
call $S_{\text{inv}}$ manifestly gauge invariant. The meaning of this gauge
invariance is left for future studies. Please note that Eq.~\eqref{eq:(5.13)}
is valid in the presence of additional interaction parameters in the action. In
deriving Eq.~\eqref{eq:(5.13)} we have only assumed that the ghost part is
given by~Eq.~\eqref{eq:(5.11)}; as long as the action satisfies the BRST
invariance~\eqref{eq:(4.11)}, we can derive Eq.~\eqref{eq:(5.13)}.

\subsection{First order}
\label{sec:5.3}
Since the gauge coupling~$e$ in~Eq.~\eqref{eq:(5.3)} accompanies the gauge
field, the first order term $S^{(1)}$ must have the structure:
\begin{equation}
   S^{(1)}=\int_{p,k}\,\Bar{\psi}(-p-k)V_\mu(p,k)\psi(p)A_\mu(k).
\label{eq:(5.18)}
\end{equation}
Because of the charge conjugation symmetry of our formulation, we can exclude
the term cubic in the gauge potential~\cite{Miyakawa:2021hcx}.

Equation~\eqref{eq:(5.3)} gives
\begin{align}
   &\left(\frac{4-D}{2}+m\frac{\partial}{\partial m}\right)S^{(1)}
\notag\\
   &=\int_k\,
   \left[
   \left(
   2k^2+\frac{D+2}{2}+k\cdot\frac{\partial}{\partial k}\right)
   A_\mu(k)
   +2(2k^2+1)
   \frac{\delta S^{(0)}}{\delta A_\mu(-k)}
   \right]
   \frac{\delta S^{(1)}}{\delta A_\mu(k)}
\notag\\
   &\qquad{}
   +\int_p\,
   S^{(1)}\frac{\overleftarrow{\delta}}{\delta\psi(p)}
   \left[
   \left(
   2p^2+\frac{D+1}{2}+p\cdot\frac{\partial}{\partial p}\right)
   \psi(p)
   +i(4p^2+1)\frac{\overrightarrow{\delta}}{\delta\Bar{\psi}(-p)}S^{(0)}
   \right]
\notag\\
   &\qquad{}
   +\int_p\,
   \left[
   \left(
   2p^2+\frac{D+1}{2}
   +p\cdot\frac{\partial}{\partial p}\right)
   \Bar{\psi}(-p)
   +i(4p^2+1)S^{(0)}\frac{\overleftarrow{\delta}}{\delta\psi(p)}
   \right]
   \frac{\overrightarrow{\delta}}{\delta\Bar{\psi}(-p)}S^{(1)}
\notag\\
   &\qquad{}
   +4\int_{p,k}\,
   \Tr\left\{
   p_\mu\left[\psi(p)+i\frac{\delta}{\delta\Bar{\psi}(-p)}S^{(0)}\right]
   S^{(0)}\frac{\overleftarrow{\delta}}{\delta\psi(p+k)}
   \right\}
   \left[A_\mu(k)+\frac{\delta S^{(0)}}{\delta A_\mu(-k)}\right]
\notag\\
   &\qquad{}
   +4\int_{p,k}\,
   \Tr\left\{
   \frac{\overrightarrow{\delta}}{\delta\Bar{\psi}(-p)}S^{(0)}
   \left[
   \Bar{\psi}(-p-k)
   +iS^{(0)}\frac{\overleftarrow{\delta}}{\delta\psi(p+k)}
   \right]
   (p+k)_\mu
   \right\}
\notag\\
   &\qquad\qquad\qquad\qquad{}
   \times\left[A_\mu(k)+\frac{\delta S^{(0)}}{\delta A_\mu(-k)}\right].
\label{eq:(5.19)}
\end{align}

At this stage, it is very helpful to introduce new variables by\footnote{Note
that $e^{k^2}\mathcal{A}_\mu(k)$, $e^{p^2}\Psi(p)$, and~$e^{p^2}\Bar{\Psi}(p)$ are
the variables of the 1PI action in the lowest order in perturbation theory (see
Eq.~(23) of~Ref.~\cite{Igarashi:2016qdr}, for example). We thus expect that
interaction vertices simplify if expressed in terms of these variables.}
\begin{subequations}\label{eq:(5.20)}
\begin{align}
   \mathcal{A}_\mu(k)
   &\equiv A_\mu(k)+\frac{\delta S^{(0)}}{\delta A_\mu(-k)}
   =e^{-2k^2}h_{\mu\nu}(k)A_\nu(k),
\\
   \Psi(p)
   &\equiv\psi(p)+i\frac{\overrightarrow{\delta}}{\delta\Bar{\psi}(-p)}S^{(0)}
   =e^{-2p^2}\frac{1}{i}h_F(p)\psi(p),
\\
   \Bar{\Psi}(-p)
   &\equiv\Bar{\psi}(-p)+iS^{(0)}\frac{\overleftarrow{\delta}}{\delta\psi(p)}
   =\Bar{\psi}(-p)e^{-2p^2}\frac{1}{i}h_F(p),
\end{align}
\end{subequations}
where
\begin{subequations}\label{eq:(5.21)}
\begin{align}
   h_{\mu\nu}(k)
   &\equiv
   \left(\delta_{\mu\nu}-\frac{k_\mu k_\nu}{k^2}\right)\frac{1}{e^{-2k^2}+k^2}
   +\frac{k_\mu k_\nu}{k^2}\frac{\xi}{\xi e^{-2k^2}+k^2},
\label{eq:(5.21a)}\\
   h_F(p)&\equiv
   \frac{i}{e^{-2p^2}+i(\Slash{p}+im)},
\end{align}
\end{subequations}
from~Eq.~\eqref{eq:(5.8)}. These $h$-functions are the high-momentum
propagators satisfying
\begin{subequations}\label{eq:(5.22)}
\begin{align}
   \left(
   k\cdot\frac{\partial}{\partial k}+2
   \right)h_{\mu\nu}(k)
   &=2(2k^2+1)e^{-2k^2}h_{\mu\rho}(k)h_{\rho\nu}(k),
\label{eq:(5.22a)}\\
   \left(
   p\cdot\frac{\partial}{\partial p}
   +m\frac{\partial}{\partial m}
   +1
   \right)h_F(p)
   &=(4p^2+1)e^{-2p^2}\frac{1}{i} h_F(p)^2,
\label{eq:(5.22b)}
\end{align}
\end{subequations}
and
\begin{equation}
   (\Slash{p}+im)h_F(p)=ie^{-2p^2}h_F(p)+1.
\label{eq:(5.23)}
\end{equation}

Using Eq.~\eqref{eq:(5.20)}, it is straightforward to show
\begin{align}
   &\left(
   k\cdot\frac{\partial}{\partial k}+\frac{D+2}{2}\right)
   \left[e^{k^2}\mathcal{A}_\mu(k)\right]
   \cdot\frac{\delta}{\delta\left[e^{k^2}\mathcal{A}_\mu(k)\right]}
\notag\\
   &=\left[
   \left(
   2k^2+\frac{D+2}{2}+k\cdot\frac{\partial}{\partial k}\right)
   A_\mu(k)
   +2(2k^2+1)
   \frac{\delta S^{(0)}}{\delta A_\mu(-k)}
   \right]
   \cdot\frac{\delta}{\delta A_\mu(k)},
\label{eq:(5.24)}\\
   &\frac{\overleftarrow{\delta}}{\delta\left[e^{p^2}\Psi(p)\right]}
   \left(
   p\cdot\frac{\partial}{\partial p}+m\frac{\partial}{\partial m}
   +\frac{D+1}{2}\right)
   \left[e^{p^2}\Psi(p)\right]
\notag\\
   &=\frac{\overleftarrow{\delta}}{\delta\psi(p)}
   \left[
   \left(
   2p^2+\frac{D+1}{2}+p\cdot\frac{\partial}{\partial p}\right)
   \psi(p)
   +i(4p^2+1)\frac{\overrightarrow{\delta}}{\delta\Bar{\psi}(-p)}S^{(0)}
   \right],
\label{eq:(5.25)}
\end{align}
and
\begin{align}
   &\left(
   p\cdot\frac{\partial}{\partial p}+m\frac{\partial}{\partial m}
   +\frac{D+1}{2}\right)\left[\Bar{\Psi}(-p)e^{p^2}\right]\cdot
   \frac{\overrightarrow{\delta}}{\delta\left[\Bar{\Psi}(-p)e^{p^2}\right]}
\notag\\
   &=\left[
   \left(
   2p^2+\frac{D+1}{2}
   +p\cdot\frac{\partial}{\partial p}\right)
   \Bar{\psi}(-p)
   +i(4p^2+1)S^{(0)}\frac{\overleftarrow{\delta}}{\delta\psi(p)}
   \right]
   \cdot\frac{\overrightarrow{\delta}}{\delta\Bar{\psi}(-p)}.
\label{eq:(5.26)}
\end{align}
In terms of these new variables, Eq.~\eqref{eq:(5.19)} becomes quite simple:
\begin{align}
   &\left(\frac{4-D}{2}+m\frac{\partial}{\partial m}\right)S^{(1)}
\notag\\
   &=\int_k\,
   \left(
   k\cdot\frac{\partial}{\partial k}+\frac{D+2}{2}\right)
   \left[e^{k^2}\mathcal{A}_\mu(k)\right]
   \cdot\frac{\delta S^{(1)}}{\delta\left[e^{k^2}\mathcal{A}_\mu(k)\right]}
\notag\\
   &\qquad{}
   +\int_p\,
   S^{(1)}\frac{\overleftarrow{\delta}}{\delta\left[e^{p^2}\Psi(p)\right]}
   \left(
   p\cdot\frac{\partial}{\partial p}+m\frac{\partial}{\partial m}
   +\frac{D+1}{2}\right)
   \left[e^{p^2}\Psi(p)\right]
\notag\\
   &\qquad{}
   +\int_p\,
   \left(
   p\cdot\frac{\partial}{\partial p}+m\frac{\partial}{\partial m}
   +\frac{D+1}{2}\right)
   \left[\Bar{\Psi}(-p)e^{p^2}\right]
   \cdot
   \frac{\overrightarrow{\delta}}{\delta\left[\Bar{\Psi}(-p)e^{p^2}\right]}
   S^{(1)}
\notag\\
   &\qquad{}
   +4\int_{p,k}\,
   p_\mu\Bar{\Psi}(-p-k)e^{2(p+k)^2}(\Slash{p}+\Slash{k}+im)\Psi(p)
   \mathcal{A}_\mu(k)
\notag\\
   &\qquad{}
   +4\int_{p,k}\,
   \Bar{\Psi}(-p-k)(\Slash{p}+im)e^{2p^2}\Psi(p)
   (p+k)_\mu
   \mathcal{A}_\mu(k).
\label{eq:(5.27)}
\end{align}
We now write
\begin{equation}
   S^{(1)}
   =\int_{p,k}\,\Bar{\Psi}(-p-k)e^{(p+k)^2}
   \widetilde{V}_\mu(p,k)e^{p^2}\Psi(p)e^{k^2}\mathcal{A}_\mu(k),
\label{eq:(5.28)}
\end{equation}
so that the vertex part~$\widetilde{V}_\mu$ satisfies the inhomogeneous scaling
equation
\begin{align}
   &\left(
   p\cdot\frac{\partial}{\partial p}
   +k\cdot\frac{\partial}{\partial k}
   +m\frac{\partial}{\partial m}
   \right)\widetilde{V}_\mu(p,k)
\notag\\
   &=4e^{(p+k)^2-p^2-k^2}(\Slash{p}+\Slash{k}+im)p_\mu
   +4e^{p^2-(p+k)^2-k^2}(\Slash{p}+im)(p+k)_\mu.
\label{eq:(5.29)}
\end{align}
We wish to find a local solution which can be expanded in powers of $p$ and~$k$
at zero momenta. Equation~\eqref{eq:(5.29)} determines $\widetilde{V}_\mu(p,k)$
up to a constant vector. A particular solution is obtained by the formula
in~Appendix~\ref{sec:A}. The general solution is 
\begin{align}
   &\widetilde{V}_\mu(p,k)
\notag\\
   &=\widetilde{V}_\mu
   +2(\Slash{p}+\Slash{k}+im)p_\mu F((p+k)^2-p^2-k^2)
   +2(\Slash{p}+im)(p+k)_\mu F(p^2-(p+k)^2-k^2),
\label{eq:(5.30)}
\end{align}
where $\Tilde{V}_\mu$ is a constant vector, and 
\begin{equation}
   F(x)\equiv\frac{e^x-1}{x}.
\label{eq:(5.31)}
\end{equation}
$\widetilde{V}_\mu$ is determined by imposing the WT
identity~\eqref{eq:(5.13)}, which requires
\begin{equation}
   k_\mu\widetilde{V}_\mu(p,k)
   =e^{(p+k)^2-p^2-k^2}(\Slash{p}+\Slash{k}+im)
   +e^{p^2-(p+k)^2-k^2}(\Slash{p}+im).
\label{eq:(5.32)}
\end{equation}
This gives $\widetilde{V}_\mu=\gamma_\mu$, and we obtain
\begin{align}
   &\widetilde{V}_\mu(p,k)
\notag\\
   &=\gamma_\mu
   +2(\Slash{p}+\Slash{k}+im)p_\mu F((p+k)^2-p^2-k^2)
   +2(\Slash{p}+im)(p+k)_\mu F(p^2-(p+k)^2-k^2).
\label{eq:(5.33)}
\end{align}
It follows from this that 
\begin{equation}
   \widetilde{V}_\mu(p+k,-k)=\widetilde{V}_\mu(p,k),
\label{eq:(5.34)}
\end{equation}
which will be used frequently below.

Our result for $S^{(1)}$ coincides with the first order term of the gauge
invariant local Wilson action obtained in~Ref.~\cite{Miyakawa:2021hcx}.

\subsection{Second order}
\label{sec:5.4}
We expect that the anomalous dimensions are second order in $e$:
\begin{equation}
   \gamma=O(e^2),\qquad
   \gamma_F=O(e^2),\qquad
   \beta_m=O(e^2).
\label{eq:(5.35)}
\end{equation}
In what follows we denote
\begin{equation}
   \gamma\equiv\gamma_1e^2+\dotsb,\qquad
   \gamma_F\equiv\gamma_{F1}e^2+\dotsb,\qquad
   \beta_m\equiv\beta_{m1}e^2+\dotsb.
\label{eq:(5.36)}
\end{equation}

Extracting the second order terms in~Eq.~\eqref{eq:(5.3)} is already a
laborious task. We obtain
\begin{align}
   &(4-D)S^{(2)}
   +m\frac{\partial}{\partial m}S^{(2)}
   +2\gamma_1\xi\frac{\partial}{\partial\xi}S^{(0)}
   +\beta_{m1}m\frac{\partial}{\partial m}S^{(0)}
\notag\\
   &\qquad{}
   +\gamma_1\int_k\,
   \mathcal{A}_\mu(k)
   \frac{\delta S^{(0)}}{\delta A_\mu(k)}
   +\gamma_{F1}
   \int_p\,
   \left[
   S^{(0)}\frac{\overleftarrow{\delta}}{\delta\psi(p)}\Psi(p)
   +\Bar{\Psi}(-p)\frac{\overrightarrow{\delta}}{\delta\Bar{\psi}(-p)}S^{(0)}
   \right]
\notag\\
   &=\int_k\,
   \left(
   k\cdot\frac{\partial}{\partial k}
   +\frac{D+2}{2}
   \right)
   \left[e^{k^2}\mathcal{A}_\mu(k)\right]
   \cdot\frac{\delta S^{(2)}}{\delta\left[e^{k^2}\mathcal{A}_\mu(k)\right]}
\notag\\
   &\qquad{}
   +\int_k\,
   (2k^2+1)
   \frac{\delta^2S^{(2)}}{\delta A_\mu(k)\delta A_\mu(-k)}
\notag\\
   &\qquad{}
   +\int_p\,
   S^{(2)}\frac{\overleftarrow{\delta}}{\delta\left[e^{p^2}\Psi(p)\right]}
   \left(
   p\cdot\frac{\partial}{\partial p}+m\frac{\partial}{\partial m}
   +\frac{D+1}{2}\right)
   \left[e^{p^2}\Psi(p)\right]
\notag\\
   &\qquad{}
   +\int_p\,
   \left(
   p\cdot\frac{\partial}{\partial p}+m\frac{\partial}{\partial m}
   +\frac{D+1}{2}\right)
   \left[e^{p^2}\Bar{\Psi}(-p)\right]
   \cdot\frac{\overrightarrow{\delta}}{\delta\left[e^{p^2}\Bar{\Psi}(-p)\right]}
   S^{(2)}
\notag\\
   &\qquad{}
   +\int_p\,
   (-i)(4p^2+1)
   \Tr\left[
   \frac{\overrightarrow{\delta}}{\delta\Bar{\psi}(-p)}
   S^{(2)}\frac{\overleftarrow{\delta}}{\delta\psi(p)}
   \right]
\notag\\
   &\qquad{}
   +\int_p\,
   i(4p^2+1)
   S^{(1)}\frac{\overleftarrow{\delta}}{\delta\psi(p)}
   \frac{\overrightarrow{\delta}}{\delta\Bar{\psi}(-p)}S^{(1)}
\notag\\
   &\qquad{}
   +\int d^Dx\,
   4i
   \mathcal{A}_\mu(x)
   S^{(1)}\frac{\overleftarrow{\delta}}{\delta\psi(x)}
   \partial_\mu\Psi(x)
\notag\\
   &\qquad{}
   +\int d^Dx\,
   (-4i)
   \mathcal{A}_\mu(x)
   \partial_\mu\Bar{\Psi}(x)
   \cdot\frac{\overrightarrow{\delta}}{\delta\Bar{\psi}(x)}S^{(1)}
\notag\\
   &\qquad{}
   +\int d^Dx\,
   (-4)
   \mathcal{A}_\mu(x)
   S^{(0)}\frac{\overleftarrow{\delta}}{\delta\psi(x)}
   \partial_\mu\frac{\overrightarrow{\delta}}{\delta\Bar{\psi}(x)}S^{(1)}
\notag\\
   &\qquad{}
   +\int d^Dx\,
   4
   \mathcal{A}_\mu(x)
   S^{(1)}\partial_\mu\frac{\overleftarrow{\delta}}{\delta\psi(x)}
   \cdot\frac{\overrightarrow{\delta}}{\delta\Bar{\psi}(x)}S^{(0)}
\notag\\
   &\qquad{}
   +\int d^Dx\,
   2\mathcal{A}_\mu(x)\mathcal{A}_\mu(x)
   S^{(0)}\frac{\overleftarrow{\delta}}{\delta\psi(x)}
   \Psi(x)
\notag\\
   &\qquad{}
   +\int d^Dx\,
   2\mathcal{A}_\mu(x)\mathcal{A}_\mu(x)
   \Bar{\Psi}(x)
   \frac{\overrightarrow{\delta}}{\delta\Bar{\psi}(x)}S^{(0)}
\notag\\
   &\qquad{}
   +\int_k\,
   (2k^2+1)
   \frac{\delta S^{(1)}}{\delta A_\mu(k)}
   \frac{\delta S^{(1)}}{\delta A_\mu(-k)}
\notag\\
   &\qquad{}
   +\int d^Dx\,
   4i
   S^{(0)}\frac{\overleftarrow{\delta}}{\delta\psi(x)}
   \partial_\mu\Psi(x)
   \cdot\frac{\delta S^{(1)}}{\delta A_\mu(x)}
\notag\\
   &\qquad{}
   +\int d^Dx\,
   (-4i)
   \partial_\mu\Bar{\Psi}(x)
   \cdot\frac{\overrightarrow{\delta}}{\delta\Bar{\psi}(x)}S^{(0)}\cdot
   \frac{\delta S^{(1)}}{\delta A_\mu(x)}
\notag\\
   &\qquad{}
   +\int d^Dx\,
   4\mathcal{A}_\mu(x)\Tr\left[
   \partial_\mu
   \frac{\overrightarrow{\delta}}{\delta\Bar{\psi}(x)}
   S^{(1)}\cdot\frac{\overleftarrow{\delta}}{\delta\psi(x)}
   \right]
\notag\\
   &\qquad{}
   +\int d^Dx\,
   (-4)\mathcal{A}_\mu(x)\Tr\left[
   \frac{\overrightarrow{\delta}}{\delta\Bar{\psi}(x)}S^{(1)}
   \partial_\mu\frac{\overleftarrow{\delta}}{\delta\psi(x)}
   \right]
\notag\\
   &\qquad{}
   +\int d^Dx\,
   (-4i)
   \mathcal{A}_\mu(x)\mathcal{A}_\mu(x)
   \Tr
   \left[
   \frac{\overrightarrow{\delta}}{\delta\Bar{\psi}(x)}
   S^{(0)}\frac{\overleftarrow{\delta}}{\delta\psi(x')}
   \right]
\notag\\
   &\qquad{}
   +\int d^Dx\,
   4i
   \frac{\delta S^{(1)}}{\delta A_\mu(x)}
   \frac{\overleftarrow{\delta}}{\delta\psi(x)}
   \partial_\mu\Psi(x)
\notag\\
   &\qquad{}
   +\int d^Dx\,
   (-4i)
   \partial_\mu\Bar{\Psi}(x)
   \cdot\frac{\overrightarrow{\delta}}{\delta\Bar{\psi}(x)}
   \frac{\delta S^{(1)}}{\delta A_\mu(x)}
\notag\\
   &\qquad{}
   +\int d^Dx\,
   (-4)
   S^{(0)}\frac{\overleftarrow{\delta}}{\delta\psi(x)}
   \partial_\mu\frac{\overrightarrow{\delta}}{\delta\Bar{\psi}(x)}
   \cdot\frac{\delta S^{(1)}}{\delta A_\mu(x)}
\notag\\
   &\qquad{}
   +\int d^Dx\,
   4
   \frac{\delta S^{(1)}}{\delta A_\mu(x)}
   \partial_\mu\frac{\overleftarrow{\delta}}{\delta\psi(x)}
   \cdot\frac{\overrightarrow{\delta}}{\delta\Bar{\psi}(x)}S^{(0)}
\notag\\
   &\qquad{}
   +\int d^Dx\,
   2\frac{\delta^2S^{(0)}}{\delta A_\mu(x)\delta A_\mu(x')}
   S^{(0)}\frac{\overleftarrow{\delta}}{\delta\psi(x)}
   \Psi(x)
\notag\\
   &\qquad{}
   +\int d^Dx\,
   2
   \frac{\delta^2S^{(0)}}{\delta A_\mu(x)\delta A_\mu(x')}
   \Bar{\Psi}(x)
   \frac{\overrightarrow{\delta}}{\delta\Bar{\psi}(x)}S^{(0)}
\notag\\
   &\qquad{}
   +4\int d^Dx\,
   \frac{\delta S^{(1)}}{\delta A_\mu(x)}
   \Tr\left[
   \partial_\mu\frac{\overrightarrow{\delta}}{\delta\Bar{\psi}(x)}S^{(0)}
   \cdot\frac{\overleftarrow{\delta}}{\delta\psi(x')}
   -\frac{\overrightarrow{\delta}}{\delta\Bar{\psi}(x')}S^{(0)}
   \partial_\mu\frac{\overleftarrow{\delta}}{\delta\psi(x)}\right],
\label{eq:(5.37)}
\end{align}
where we have used variables defined in~Eq.~\eqref{eq:(5.20)}. Note that we
take the limit $x'\to x$ only after taking differentials as has been explained
in~Sect.~\ref{sec:3}. In Appendix~\ref{sec:B} we elaborate on how this limit
actually works in this case.

We have four types of terms:
\begin{equation}
   S^{(2)}
   =S^{(2)}\bigr|_{\Bar{\psi}AA\psi}
   +S^{(2)}\bigr|_{\Bar{\psi}\psi\Bar{\psi}\psi}
   +S^{(2)}\bigr|_{AA}
   +S^{(2)}\bigr|_{\Bar{\psi}\psi}.
\label{eq:(5.38)}
\end{equation}
We compute them one by one.

\subsection{$\Bar{\psi}AA\psi$ term}
\label{sec:5.5}
Let us first consider the term proportional to~$\Bar{\psi}AA\psi$:
\begin{align}
   S^{(2)}|_{\Bar{\psi}AA\psi}
   \equiv\int_{p,k,l}\Bar{\Psi}(-p-k-l)e^{(p+k+l)^2}\widetilde{V}_{\mu\nu}(p,k,l)
   e^{p^2}\Psi(p)e^{k^2}\mathcal{A}_\mu(k)e^{l^2}\mathcal{A}_\nu(l).
\label{eq:(5.39)}
\end{align}
Equation~\eqref{eq:(5.37)} gives
\begin{align}
   &\left(
   p\cdot\frac{\partial}{\partial p}
   +k\cdot\frac{\partial}{\partial k}
   +l\cdot\frac{\partial}{\partial l}
   +m\frac{\partial}{\partial m}
   +1
   \right)\widetilde{V}_{\mu\nu}(p,k,l)
\notag\\
   &=\widetilde{V}_\mu(p+l,k)(-1)[4(p+l)^2+1]e^{-2(p+l)^2}h_F(p+l)^2
   \widetilde{V}_\nu(p,l)
\notag\\
   &\qquad{}
   +4\widetilde{V}_\mu(p+l,k)h_F(p+l)
   \Bigl[
   e^{(p+l)^2-p^2-l^2}(\Slash{p}+\Slash{l}+im)p_\nu
\notag\\
   &\qquad\qquad\qquad\qquad\qquad\qquad\qquad{}
   +e^{p^2-(p+l)^2-l^2}(\Slash{p}+im)(p+l)_\nu
   \Bigr]
\notag\\
   &\qquad{}
   +4\Bigl[
   e^{(p+k+l)^2-(p+l)^2-k^2}(\Slash{p}+\Slash{k}+\Slash{l}+im)(p+l)_\mu
\notag\\
   &\qquad\qquad\qquad\qquad\qquad{}
   +e^{(p+l)^2-(p+k+l)^2-k^2}(\Slash{p}+\Slash{l}+im)(p+k+l)_\mu
   \Bigr]
   h_F(p+l)\widetilde{V}_\nu(p,l)
\notag\\
   &\qquad{}
   -4\left[
   e^{(p+l)^2-p^2-l^2}\widetilde{V}_\mu(p+l,k)p_\nu
   +e^{(p+l)^2-(p+k+l)^2-k^2}(p+k+l)_\mu\widetilde{V}_\nu(p,l)
   \right]
\notag\\
   &\qquad{}
   -2\delta_{\mu\nu}e^{-k^2-l^2}
   \left[
   e^{(p+k+l)^2-p^2}(\Slash{p}+\Slash{k}+\Slash{l}+im)
   +e^{p^2-(p+k+l)^2}(\Slash{p}+im)
   \right],
\label{eq:(5.40)}
\end{align}
where we have used the relation~\eqref{eq:(5.23)}. Noting further the
properties~\eqref{eq:(5.22)}, we can simplify the above to
\begin{align}
   &\left(
   p\cdot\frac{\partial}{\partial p}
   +k\cdot\frac{\partial}{\partial k}
   +l\cdot\frac{\partial}{\partial l}
   +m\frac{\partial}{\partial m}
   +1
   \right)\widetilde{V}_{\mu\nu}(p,k,l)
\notag\\
   &=\left(
   p\cdot\frac{\partial}{\partial p}
   +k\cdot\frac{\partial}{\partial k}
   +l\cdot\frac{\partial}{\partial l}
   +m\frac{\partial}{\partial m}
   +1
   \right)
   \widetilde{V}_\mu(p+l,k)h_F(p+l)
   \widetilde{V}_\nu(p,l)
\notag\\
   &\qquad{}
   -4\left[
   e^{(p+l)^2-p^2-l^2}\widetilde{V}_\mu(p+l,k)p_\nu
   +e^{(p+l)^2-(p+k+l)^2-k^2}(p+k+l)_\mu\widetilde{V}_\nu(p,l)
   \right]
\notag\\
   &\qquad{}
   -2\delta_{\mu\nu}e^{-k^2-l^2}
   \left[
   e^{(p+k+l)^2-p^2}(\Slash{p}+\Slash{k}+\Slash{l}+im)
   +e^{p^2-(p+k+l)^2}(\Slash{p}+im)
   \right].
\label{eq:(5.41)}
\end{align}
The last line can be integrated by the formula in~Appendix~\ref{sec:A} as
\begin{align}
   &-2\delta_{\mu\nu}e^{-k^2-l^2}
   \left[
   e^{(p+k+l)^2-p^2}(\Slash{p}+\Slash{k}+\Slash{l}+im)
   +e^{p^2-(p+k+l)^2}(\Slash{p}+im)
   \right]
\notag\\
   &=-\delta_{\mu\nu}\left(
   p\cdot\frac{\partial}{\partial p}
   +k\cdot\frac{\partial}{\partial k}
   +l\cdot\frac{\partial}{\partial l}
   +m\frac{\partial}{\partial m}
   +1
   \right)
\notag\\
   &\qquad{}
   \times
   \bigl[
   (\Slash{p}+\Slash{k}+\Slash{l}+im)
   F((p+k+l)^2-p^2-k^2-l^2)
\notag\\
   &\qquad\qquad\qquad\qquad{}
   +(\Slash{p}+im)
   F(p^2-(p+k+l)^2-k^2-l^2)
   \bigr],
\label{eq:(5.42)}
\end{align}
where the function~$F(x)$ is defined by~Eq.~\eqref{eq:(5.31)}. Therefore, the
solution to~Eq.~\eqref{eq:(5.41)} is given by
\begin{align}
   \widetilde{V}_{\mu\nu}(p,k,l)
   &=\widetilde{V}_\mu(p+l,k)h_F(p+l)\widetilde{V}_\nu(p,l)
\notag\\
   &\qquad{}
   -\delta_{\mu\nu}
   \bigl[
   (\Slash{p}+\Slash{k}+\Slash{l}+im)
   F((p+k+l)^2-p^2-k^2-l^2)
\notag\\
   &\qquad\qquad\qquad\qquad{}
   +(\Slash{p}+im)
   F(p^2-(p+k+l)^2-k^2-l^2)
   \bigr]
\notag\\
   &\qquad{}
   -4X_{\mu\nu}(p,k,l),
\label{eq:(5.43)}
\end{align}
where $X_{\mu\nu}(p,k,l)$ satisfies
\begin{align}
   &\left(
   p\cdot\frac{\partial}{\partial p}
   +k\cdot\frac{\partial}{\partial k}
   +l\cdot\frac{\partial}{\partial l}
   +m\frac{\partial}{\partial m}
   +1
   \right)X_{\mu\nu}(p,k,l)
\notag\\
   &=e^{(p+l)^2-p^2-l^2}\widetilde{V}_\mu(p+l,k)p_\nu
   +e^{(p+l)^2-(p+k+l)^2-k^2}(p+k+l)_\mu\widetilde{V}_\nu(p,l)
\notag\\
   &=e^{(p+l)^2-p^2-l^2}
\notag\\
   &\qquad{}
   \times
   \bigl[\gamma_\mu
   +2(\Slash{p}+\Slash{k}+\Slash{l}+im)(p+l)_\mu F((p+k+l)^2-(p+l)^2-k^2)
\notag\\
   &\qquad\qquad{}
   +2(\Slash{p}+\Slash{l}+im)(p+k+l)_\mu F((p+l)^2-(p+k+l)^2-k^2)\bigr]
   p_\nu
\notag\\
   &\qquad{}
   +e^{(p+l)^2-(p+k+l)^2-k^2}(p+k+l)_\mu
\notag\\
   &\qquad\qquad{}
   \times
   \bigl[\gamma_\nu
   +2(\Slash{p}+\Slash{l}+im)p_\nu F((p+l)^2-p^2-l^2)
\notag\\
   &\qquad\qquad\qquad{}
   +2(\Slash{p}+im)(p+l)_\nu F(p^2-(p+l)^2-l^2)\bigr].
\label{eq:(5.44)}
\end{align}
This can be solved again by the formula in~Appendix~\ref{sec:A} to yield
\begin{align}
   &X_{\mu\nu}(p,k,l)
\notag\\
   &=\frac{1}{2}\gamma_\mu p_\nu F((p+l)^2-p^2-l^2)
   +\frac{1}{2}(p+k+l)_\mu\gamma_\nu F((p+l)^2-(p+k+l)^2-k^2)
\notag\\
   &\qquad{}
   +(\Slash{p}+\Slash{l}+im)(p+k+l)_\mu p_\nu
   F((p+l)^2-(p+k+l)^2-k^2)F((p+l)^2-p^2-l^2)
\notag\\
   &\qquad{}
   +\frac{(\Slash{p}+\Slash{k}+\Slash{l}+im)(p+l)_\mu p_\nu}
   {(p+k+l)^2-(p+l)^2-k^2}
\notag\\
   &\qquad\qquad{}
   \times
   \left[F((p+k+l)^2-p^2-k^2-l^2)-F((p+l)^2-p^2-l^2)\right]
\notag\\
   &\qquad{}
   +\frac{(\Slash{p}+im)(p+k+l)_\mu(p+l)_\nu}{p^2-(p+l)^2-l^2}
\notag\\
   &\qquad\qquad{}
   \times
   \left[F(p^2-(p+k+l)^2-k^2-l^2)-F((p+l)^2-(p+k+l)^2-k^2)\right],
\label{eq:(5.45)}
\end{align}
where we have used the identity
\begin{equation}
   F(x)F(y)=\left(\frac{1}{x}+\frac{1}{y}\right)F(x+y)
   -\frac{F(x)}{y}-\frac{F(y)}{x}.
\label{eq:(5.46)}
\end{equation}
Equation~\eqref{eq:(5.43)} with $X_{\mu\nu}$ given by Eq.~\eqref{eq:(5.45)}
gives a local solution to~Eq.~\eqref{eq:(5.40)}. The solution is unique because
the homogeneous equation
\begin{equation}
   \left(
   p\cdot\frac{\partial}{\partial p}
   +k\cdot\frac{\partial}{\partial k}
   +l\cdot\frac{\partial}{\partial l}
   +m\frac{\partial}{\partial m}
   +1
   \right)\widetilde{V}_{\mu\nu}(p,k,l)=0
\label{eq:(5.47)}
\end{equation}
has no solution analytic in momenta and~$m$.

\subsection{$\Bar{\psi}\psi\Bar{\psi}\psi$ term}
\label{sec:5.6}
We observe that the inhomogeneous terms of~Eq.~\eqref{eq:(5.37)} that can
contribute to the four-Fermi term always contain the
factor~$\delta S^{(1)}/\delta A_\mu(x)$, where $S^{(1)}$ is given
by~Eq.~\eqref{eq:(5.18)}. This suggests the structure
\begin{align}
   &S^{(2)}|_{\Bar{\psi}\psi\Bar{\psi}\psi}
\notag\\
   &=\frac{1}{2}\int_{p,q,k}
   \Bar{\Psi}(-p-k)e^{(p+k)^2}\Gamma_\mu(p,k)e^{p^2}\Psi(p)
   \Bar{\Psi}(-q)e^{q^2}\Gamma_\nu(q+k,-k)e^{(q+k)^2}\Psi(q+k)D_{\mu\nu}(k).
\label{eq:(5.48)}
\end{align}
Using Eq.~\eqref{eq:(5.29)} and the properties~\eqref{eq:(5.22)}, we find that
the ERG equation for~Eq.~\eqref{eq:(5.48)} takes the following extremely simple
form:
\begin{align}
   &\left(
   p\cdot\frac{\partial}{\partial p}
   +q\cdot\frac{\partial}{\partial q}
   +k\cdot\frac{\partial}{\partial k}
   +m\frac{\partial}{\partial m}
   +2
   \right)
   \left[
   \Gamma_\mu(p,k)\cdot\Gamma_\nu(q+k,k)D_{\mu\nu}(k)\right]   
\notag\\
   &=\left(
   p\cdot\frac{\partial}{\partial p}
   +q\cdot\frac{\partial}{\partial q}
   +k\cdot\frac{\partial}{\partial k}
   +m\frac{\partial}{\partial m}
   +2
   \right)
   \left[
   \widetilde{V}_\mu(p,k)\cdot\widetilde{V}_\nu(q+k,k)h_{\mu\nu}(k)\right].
\label{eq:(5.49)}
\end{align}
Since the corresponding homogeneous equation
\begin{equation}
   \left(
   p\cdot\frac{\partial}{\partial p}
   +q\cdot\frac{\partial}{\partial q}
   +k\cdot\frac{\partial}{\partial k}
   +m\frac{\partial}{\partial m}
   +2
   \right)
   \left[
   \Gamma_\mu(p,k)\cdot\Gamma_\nu(q+k,k)D_{\mu\nu}(k)\right]=0
\label{eq:(5.50)}
\end{equation}
has no solution analytic in momenta and~$m$, we get the unique local solution
\begin{align}
   &S^{(2)}|_{\Bar{\psi}\psi\Bar{\psi}\psi}
\notag\\
   &=\frac{1}{2}\int_{p,q,k}
   \Bar{\Psi}(-p-k)e^{(p+k)^2}\widetilde{V}_\mu(p,k)e^{p^2}\Psi(p)
   \Bar{\Psi}(-q)e^{q^2}\widetilde{V}_\nu(q+k,-k)e^{(q+k)^2}\Psi(q+k)
   h_{\mu\nu}(k).
\label{eq:(5.51)}
\end{align}

\subsection{$AA$ term and $\gamma_1$}
\label{sec:5.7}
Now, we can study the second order correction to the $AA$ term. We will see
that the analyticity of this term determines the first nontrivial order
coefficient of the anomalous dimension, $\gamma_1$ in~Eq.~\eqref{eq:(5.36)}. We
first note that the WT identity~\eqref{eq:(5.13)} requires that this correction
be transverse:
\begin{equation}
   S^{(2)}|_{AA}
   \equiv\frac{1}{2}\int_k\,
   e^{k^2}\mathcal{A}_\mu(k)
   e^{k^2}\mathcal{A}_\nu(-k)
   \left(\delta_{\mu\nu}-\frac{k_\mu k_\nu}{k^2}\right)\widetilde{V}_T(k).
\label{eq:(5.52)}
\end{equation}
For this to be local, $\Tilde{V}_T (k)$ must be of order~$k^2$ at~$k=0$. We may
also normalize the kinetic term by demanding $\Tilde{V}_T (k)$ to be of
order~$(k^2)^2$.

Now, the part of the ERG equation~\eqref{eq:(5.37)} relevant to the $AA$ term
gives
\begin{align}
   &\frac{1}{2}
   \left\{
   \left[
   k\cdot\frac{\partial}{\partial k}
   +m\frac{\partial}{\partial m}
   -2+(4-D)\right]\widetilde{V}_T(k)
   -2\gamma_1k^2
   \right\}\left(\delta_{\mu\nu}-\frac{k_\mu k_\nu}{k^2}\right)
\notag\\
   &=-\int_p\,
   \Tr\left[
   \left(
   p\cdot\frac{\partial}{\partial p}
   +m\frac{\partial}{\partial m}
   +1\right)h_F(p)\cdot\widetilde{V}_{\mu\nu}(p,-k,k)   
   \right]
\notag\\
   &\qquad{}
   -4ie^{-k^2}\int_p\,
   e^{-(p+k)^2-p^2}\Tr\left[
   h_F(p+k)\widetilde{V}_\nu(p,k)h_F(p)
   \right](2p+k)_\mu
\notag\\
   &\qquad{}
   -4e^{-2k^2}\delta_{\mu\nu}
   \int_p\,e^{-2p^2}\Tr\left[
   \frac{1}{e^{-2p^2}+i(\Slash{p}+im)}
   \right]
\notag\\
   &\equiv I_{\mu\nu}(k),
\label{eq:(5.53)}
\end{align}
where we have used Eq.~\eqref{eq:(5.22)}. Since the lhs is symmetric
under~$k\leftrightarrow-k$ and~$\mu\leftrightarrow\nu$, we make this symmetry
manifest also on the rhs by rewriting
\begin{align}
   I_{\mu\nu}(k)
   &=-\frac{1}{2}\int_p\,
   \Tr\left[
   \left(
   p\cdot\frac{\partial}{\partial p}
   +m\frac{\partial}{\partial m}
   +1\right)h_F(p)\cdot
   \widetilde{V}_{\mu\nu}(p,-k,k)
   \right]
\notag\\
   &\qquad{}
   -\frac{1}{2}\int_p\,
   \Tr\left[
   \left(
   p\cdot\frac{\partial}{\partial p}
   +k\cdot\frac{\partial}{\partial k}
   +m\frac{\partial}{\partial m}
   +1\right)h_F(p+k)\cdot
   \widetilde{V}_{\nu\mu}(p+k,k,-k)
   \right]
\notag\\
   &\qquad{}
   -2ie^{-k^2}\int_p\,
   e^{-(p+k)^2-p^2}\Tr\left[
   h_F(p+k)\widetilde{V}_\nu(p,k)h_F(p)
   \right](2p+k)_\mu
\notag\\
   &\qquad{}
   -2ie^{-k^2}\int_p\,
   e^{-(p+k)^2-p^2}\Tr\left[
   h_F(p)\widetilde{V}_\mu(p,k)h_F(p+k)
   \right](2p+k)_\nu
\notag\\
   &\qquad{}
   -4e^{-2k^2}\delta_{\mu\nu}
   \int_p\,e^{-2p^2}\Tr\left[
   \frac{1}{e^{-2p^2}+i(\Slash{p}+im)}
   \right],
\label{eq:(5.54)}
\end{align}
where we have used Eq.~\eqref{eq:(5.34)}. By using Eqs.~\eqref{eq:(5.29)},
\eqref{eq:(5.43)}, \eqref{eq:(5.44)}, and
\begin{equation}
   \left(k\cdot\frac{\partial}{\partial k}+2\right)F(-2k^2)=2e^{-2k^2},
\label{eq:(5.55)}
\end{equation}
we can show
\begin{align}
   I_{\mu\nu}(k)
   &=-\frac{1}{2}\int_p\,
   \left(
   p\cdot\frac{\partial}{\partial p}
   +k\cdot\frac{\partial}{\partial k}
   +m\frac{\partial}{\partial m}
   +2\right)
\notag\\
   &\qquad\qquad{}
   \times
   \Tr\Bigl\{
   h_F(p)\widetilde{V}_\mu(p,k)h_F(p+k)\widetilde{V}_\nu(p,k)
\notag\\
   &\qquad\qquad\qquad{}
   -4\left[h_F(p)X_{\mu\nu}(p,-k,k)+h_F(p+k)X_{\nu\mu}(p+k,k,-k)\right]
\notag\\
   &\qquad\qquad\qquad{}
   -4i\delta_{\mu\nu}F(-2k^2)e^{-2p^2}h_F(p)
   \Bigr\}.
\label{eq:(5.56)}
\end{align}

Now, let us compute $\gamma_1$ for~$D=4$. The lhs of~Eq.~\eqref{eq:(5.53)}
gives
\begin{equation}
   -2\gamma_1\left(k^2\delta_{\mu\nu}-k_\mu k_\nu\right)
\label{eq:(5.57)}
\end{equation}
to order~$k^2$. We can determine $\gamma_1$ by calculating $I_{\mu\nu}(k)$ to
the same order. Since $k\cdot\partial/\partial k=2$ for the $k^2$ term,
Eq.~\eqref{eq:(5.56)} gives
\begin{align}
   &-2\gamma_1\left(\delta_{\mu\nu}k^2-k_\mu k_\nu\right)
\notag\\
   &=-\int_p\,\frac{\partial}{\partial p_\rho}
   \Bigl(p_\rho
   \Tr\Bigl\{
   h_F(p)\widetilde{V}_\mu(p,k)h_F(p+k)\widetilde{V}_\nu(p,k)
\notag\\
   &\qquad\qquad\qquad\qquad{}
   -4\left[h_F(p)X_{\mu\nu}(p,-k,k)+h_F(p+k)X_{\nu\mu}(p+k,k,-k)\right]
\notag\\
   &\qquad\qquad\qquad\qquad{}
   -4i\delta_{\mu\nu}F(-2k^2)e^{-2p^2}h_F(p)
   \Bigr\}
   \Bigr)\Bigr|_{m=0,O(k^2)}.
\label{eq:(5.58)}
\end{align}
The 4-momentum integral on the rhs is thus given by a surface integral
at~$|p|\to\infty$. From the explicit form of the integrand, it is not difficult
to find the surface term that contributes to the integral at~$|p|=\infty$. For
instance, the last term does not contribute because of the factor~$e^{-2p^2}$.
In this way, we obtain
\begin{equation}
   -2\gamma_1=-\frac{1}{(4\pi)^2}\frac{8}{3}.
\label{eq:(5.59)}
\end{equation}
Hence, from~Eq.~\eqref{eq:(5.4)} the beta function of~$e^2$ is
\begin{equation}
   \beta(e^2)
   =-2\gamma(e^2)e^2
   \simeq-\frac{1}{(4\pi)^2}\frac{8}{3}(e^2)^2.
\label{eq:(5.60)}
\end{equation}
This agrees with the 1-loop beta function of QED.

\subsection{$\Bar{\psi}\psi$ term and $\gamma_{F1}$, $\beta_m$}
\label{sec:5.8}
Finally, we consider the second order correction to the fermion kinetic and
mass terms:
\begin{equation}
   S^{(2)}|_{\Bar{\psi}\psi}
   \equiv\int_p\Bar{\Psi}(-p)e^{p^2}\widetilde{V}_F(p)e^{p^2}\Psi(p).
\label{eq:(5.61)}
\end{equation}
The GFERG equation is given by
\begin{align}
   &\left[
   p\cdot\frac{\partial}{\partial p}+m\frac{\partial}{\partial m}
   -1+(4-D)
   \right]\widetilde{V}_F(p)
   -\beta_{m1}im
   -2\gamma_{F1}(\Slash{p}+im)
\notag\\
   &=\int_k\,
   \left(k\cdot\frac{\partial}{\partial k}+2\right)h_{\mu\nu}(k)\cdot
   \widetilde{V}_{\mu\nu}(p,-k,k)
\notag\\
   &\qquad{}
   +\int_k\,
   h_{\mu\nu}(k)
   \widetilde{V}_\mu(p,k)
   \left(p\cdot\frac{\partial}{\partial p}
   +k\cdot\frac{\partial}{\partial k}
   +m\frac{\partial}{\partial m}+1\right)
   h_F(p+k)\cdot
   \widetilde{V}_\nu(p,k)
\notag\\
   &\qquad{}
   +4i\int_k\,
   h_{\mu\nu}(k)e^{-(p+k)^2-p^2-k^2}h_F(p+k)p_\mu\widetilde{V}_\nu(p,k)
\notag\\
   &\qquad{}
   +4\int_k\,
   h_{\mu\nu}(k)e^{p^2-(p+k)^2-k^2}(\Slash{p}+im)(p+k)_\mu
   h_F(p+k)\widetilde{V}_\nu(p,k)
\notag\\
   &\qquad{}
   +4i\int_k\,
   h_{\mu\nu}(k)\widetilde{V}_\mu(p,k)e^{-(p+k)^2-p^2-k^2}h_F(p+k)p_\nu
\notag\\
   &\qquad{}
   +4\int_k\,
   h_{\mu\nu}(k)\widetilde{V}_\mu(p,k)e^{p^2-(p+k)^2-k^2}h_F(p+k)
   (\Slash{p}+im)(p+k)_\nu
\notag\\
   &\qquad{}
   -4\int_k\,h_{\mu\mu}(k)(\Slash{p}+im)e^{-2k^2}
\notag\\
   &\equiv I_F(p),
\label{eq:(5.62)}
\end{align}
where we have used Eq.~\eqref{eq:(5.22)} for the first two lines on the rhs. We
have also used~Eq.~\eqref{eq:(5.34)}. We may normalize the kinetic and mass
term so that $\Tilde{V}_F(p)$ has no term proportional to either $\Slash{p}$
or~$m$. By using Eqs.~\eqref{eq:(5.43)}, \eqref{eq:(5.29)}, \eqref{eq:(5.44)},
\eqref{eq:(5.23)}, and~\eqref{eq:(5.55)}, we obtain
\begin{equation}
   I_F(p)
   =\int_k\,
   \left(
   p\cdot\frac{\partial}{\partial p}
   +k\cdot\frac{\partial}{\partial k}
   +m\frac{\partial}{\partial m}
   +3
   \right)
   \left[h_{\mu\nu}(k)\widetilde{V}_{\mu\nu}(p,-k,k)\right].
\label{eq:(5.63)}
\end{equation}

Now, let us compute $\beta_{m1}$, $\gamma_{F1}$ for~$D=4$. The lhs
of~Eq.~\eqref{eq:(5.62)} gives
\begin{equation}
   -\beta_{m1}im-2\gamma_{F1}\left(\Slash{p}+im\right)
\label{eq:(5.64)}
\end{equation}
to first order in~$\Slash{p}$ and~$m$. We can determine $\beta_{m1}$
and~$\gamma_{F1}$ by calculating $I_F(p)$ to the same order. We can take
$p\cdot\partial/\partial p+m\partial/\partial m=1$, and Eqs.~\eqref{eq:(5.62)}
and~\eqref{eq:(5.63)} give
\begin{align}
   -\beta_{m1}im
   -2\gamma_{F1}(\Slash{p}+im)
   =\int_k\,\frac{\partial}{\partial k_\rho}
   \left.\left[k_\rho h_{\mu\nu}(k)\widetilde{V}_{\mu\nu}(p,-k,k)
   \right]\right|_{O(m),O(p)}.
\label{eq:(5.65)}
\end{align}
It is again straightforward to find the surface term at~$|k|=\infty$, which
contributes to this integral, and we obtain
\begin{equation}
   \beta_{m1}=\frac{6}{(4\pi)^2},\qquad
   \gamma_{F1}=\frac{3}{(4\pi)^2}.
\label{eq:(5.66)}
\end{equation}
The former is the usual mass anomalous dimension in QED. Interestingly, the
latter coincides with the anomalous dimension resulting from the wave function
renormalization of the \emph{flowed\/} or \emph{diffused\/} (i.e., not usual)
fermion field; the 1-loop renormalization factor has been given in~Eq.~(2.16)
of~Ref.~\cite{Luscher:2013cpa}, where $C_F=1$ for QED with the electron. Note
that this anomalous dimension is independent of the gauge-fixing
parameter~$\xi$. This is expected because the Wilson action~\eqref{eq:(3.1)},
by construction, reproduces the correlation functions of flowed or diffused
fields up to contact terms~\cite{Sonoda:2020vut}.

\section{Conclusion}
\label{sec:6}
In this paper we have constructed the gradient flow exact renormalization group
(GFERG) for QED, based on the BRST invariant diffusion
equations~\eqref{eq:(2.16)}. With the exclusion of the gauge fixing term, the
Wilson action~\eqref{eq:(5.16)} becomes manifestly invariant under the gauge
transformation~\eqref{eq:(5.17)}. We have computed the action perturbatively in
powers of the coupling~$e$ to the order~$e^2$, reproducing the 1-loop beta
function and anomalous dimensions. It was especially pleasing to find the
anomalous dimension of the electron field as gauge invariant.

Our perturbative calculations show that the Wilson action becomes complex
despite the simplicity in gauge invariance. The complexity comes from that of
the GFERG differential equations. But we believe that the manifest gauge
invariance will turn out to be a big advantage when we attempt to solve the
GFERG differential equations non-perturbatively (with some gauge invariant
approximations).

Whether the GFERG differential equation~\eqref{eq:(4.10)} has a non-trivial
fixed-point satisfying the WT identity~\eqref{eq:(5.13)} is of much interest to
be studied in the future. See, for example, Ref.~\cite{Igarashi:2021zml} and
references cited therein for related studies. At present we even do not know
what it means to have a fixed-point in the GFERG formalism.

The GFERG formalism was originally introduced for non-abelian gauge
theories~\cite{Sonoda:2020vut}. It should be interesting to extend the analysis
of this paper to see how far we can simplify the realization of non-abelian
gauge invariance compared with the standard ERG formalism.

\section*{Acknowledgments}
This work was partially supported by Japan Society for the Promotion of Science
(JSPS) Grant-in-Aid for Scientific Research Grant Number JP20H01903.
Hiroshi Suzuki would like to thank Katsumi Itoh for informative discussions.

\appendix

\section{Integration formula}
\label{sec:A}
A particular solution to
\begin{equation}
   \left(
   \sum_ip_i\cdot\frac{\partial}{\partial p_i}
   +m\frac{\partial}{\partial m}
   +\zeta\right)F(p,m)
   =f(p,m),
\label{eq:(A1)}
\end{equation}
where
\begin{equation}
   \lim_{\alpha\to0}\alpha^\zeta f(\alpha p,\alpha m)=0,
\label{eq:(A2)}
\end{equation}
is given by
\begin{equation}
   F(p,m)=\int_0^1d\alpha\,\alpha^{\zeta-1}f(\alpha p,\alpha m).
\label{eq:(A3)}
\end{equation}
Note that this solution is analytic in the momenta and~$m$.

The proof is straightforward. Noting
\begin{equation}
   \frac{\partial}{\partial\alpha}f(\alpha p,\alpha m)
   =\frac{1}{\alpha}
   \left(
   \sum_ip_i\cdot\frac{\partial}{\partial p_i}
   +m\frac{\partial}{\partial m}
   \right)
   f(\alpha p,\alpha m)
\label{eq:(A4)}
\end{equation}
under the prerequisite~\eqref{eq:(A2)}, we have
\begin{align}
   &\left(
   \sum_ip_i\cdot\frac{\partial}{\partial p_i}
   +m\frac{\partial}{\partial m}
   \right)
   \int_0^1d\alpha\,\alpha^{\zeta-1}f(\alpha p,\alpha m)
\notag\\
   &=\int_0^1d\alpha\,\alpha^\zeta\frac{\partial}{\partial\alpha}
   f(\alpha p,\alpha m)
\notag\\
   &=f(p,m)
   -\zeta\int_0^1d\alpha\,\alpha^{\zeta-1}f(\alpha p,\alpha m).
\label{eq:(A5)}
\end{align}
This is Eq.~\eqref{eq:(A1)} for~Eq.~\eqref{eq:(A3)}.

\section{The working of the limit $x'\to x$}
\label{sec:B}
In Sect.~\ref{sec:3.1} we have explained how to take second and higher order
functional differentials at the same point as a limit of functional
differentials at different points. This careful treatment is necessary to avoid
unphysical singularities. In deriving the GFERG differential equation for the
second order Wilson action~$S^{(2)}$ we need to practice the treatment. There
are three integrals to consider.

\subsection{$AA$ term}
We compute
\begin{align}
   &\int d^Dx\,(-4i)\mathcal{A}_\mu(x)\mathcal{A}_\mu(x)
   \Tr\left[
   \frac{\overrightarrow{\delta}}{\delta\Bar{\psi}(x)}S^{(0)}
   \frac{\overleftarrow{\delta}}{\delta\psi(x')}\right]
\notag\\
   &=\int_k\,(-4i)\mathcal{A}_\mu(k)\mathcal{A}_\mu(-k)
   \int_p\,\Tr
   (-)\frac{\Slash{p}+im}{e^{-2p^2}+i\left(\Slash{p}+im\right)}
   e^{-ip(x-x')}
\notag\\
   &=\int_k\,(-4i)\mathcal{A}_\mu(k)\mathcal{A}_\mu(-k)
   \int_p\,i\Tr
   \frac{-e^{-2p^2}+e^{-2p^2}+i\left(\Slash{p}+im\right)}
   {e^{-2p^2}+i\left(\Slash{p}+im\right)}e^{-ip(x-x')}
\notag\\
   &=-4\int_k\,\mathcal{A}_\mu(k)\mathcal{A}_\mu(-k)
   \int_p\,e^{-ip(x-x')}e^{-2p^2}
   \Tr\frac{1}{e^{-2p^2}+i\left(\Slash{p}+im\right)}
\notag\\
   &\qquad{}
   +16\int_k\,\mathcal{A}_\mu(k)\mathcal{A}_\mu(-k)
   \int_p\,e^{-ip(x-x')}.
\label{eq:(B1)}
\end{align}
For~$x\neq x'$, we find
\begin{equation}
   \int_p\,e^{-ip(x-x')}=\delta(x-x')=0.
\label{eq:(B2)}
\end{equation}
Hence, in the limit~$x'\to x$, we obtain
\begin{equation}
   -4\int_k\,\mathcal{A}_\mu(k)\mathcal{A}_\mu(-k)
   \int_p\,e^{-2p^2}\Tr\frac{1}{e^{-2p^2}+i\left(\Slash{p}+im\right)},
\label{eq:(B3)}
\end{equation}
where the integral over~$p$ is absolutely convergent.

\subsection{$\Bar{\psi}\psi$ term}
We compute
\begin{align}
   &2\int d^Dx\,
   \frac{\delta^2S^{(0)}}{\delta A_\mu(x)\delta A_\mu(x')}
   \left[
   S^{(0)}\frac{\overleftarrow{\delta}}{\delta\psi(x)}\Psi(x)
   +\Bar{\Psi}(x)\frac{\overrightarrow{\delta}}{\delta\Bar{\psi}(x)}S^{(0)}
   \right]
\notag\\
   &=2\int_k\,
   e^{ik(x-x')}
   \left[(D-1)\frac{k^2}{e^{-2k^2}+k^2}+\frac{k^2}{\xi e^{-2k^2}+k^2}\right]
\notag\\
   &\qquad{}
   \times\int_p\,
   \left[
   \Bar{\psi}(-p)\frac{\Slash{p}+im}{e^{-2p^2}+i\left(\Slash{p}+im\right)}
   \Psi(p)
   +\Bar{\Psi}(-p)\frac{\Slash{p}+im}{e^{-2p^2}+i\left(\Slash{p}+im\right)}
   \psi(p)
   \right]
\notag\\
   &=2\int_k\,
   e^{ik(x-x')}
   \left[
   D-(D-1)\frac{e^{-2k^2}}{e^{-2k^2}+k^2}
   -\frac{\xi e^{-2k^2}}{\xi e^{-2k^2}+k^2}
   \right]
\notag\\
   &\qquad{}
   \times\int_p\,
   \left[
   \Bar{\psi}(-p)\frac{\Slash{p}+im}{e^{-2p^2}+i\left(\Slash{p}+im\right)}
   \Psi(p)
   +\Bar{\Psi}(-p)\frac{\Slash{p}+im}{e^{-2p^2}+i\left(\Slash{p}+im\right)}
   \psi(p)
   \right].
\label{eq:(B4)}
\end{align}
Ignoring the delta function again, we obtain
\begin{align}
   &\stackrel{x'\to x}{\longrightarrow}
   -2\int_k\,
   e^{-2k^2}h_{\mu\mu}(k)
   \int_p\,
   \biggl[
   \Bar{\psi}(-p)\frac{\Slash{p}+im}{e^{-2p^2}+i\left(\Slash{p}+im\right)}
   \Psi(p)
\notag\\
   &\qquad\qquad\qquad\qquad\qquad\qquad\qquad{}
   +\Bar{\Psi}(-p)\frac{\Slash{p}+im}{e^{-2p^2}+i\left(\Slash{p}+im\right)}
   \psi(p)
   \biggr],
\label{eq:(B5)}
\end{align}
where $h_{\mu\nu}(k)$ is defined by~Eq.~\eqref{eq:(5.21a)}.

\subsection{Vanishing terms}
We examine the last integral of~Eq.~\eqref{eq:(5.37)}.
\begin{align}
   &4\int d^Dx\,
   \frac{\delta S^{(1)}}{\delta A_\mu(x)}
   \Tr\left[
   \partial_\mu\frac{\overrightarrow{\delta}}{\delta\Bar{\psi}(x)}
   S^{(0)}
   \frac{\overleftarrow{\delta}}{\delta\psi(x')}
   -\frac{\overrightarrow{\delta}}{\delta\Bar{\psi}(x')}
   S^{(0)}
   \partial_\mu\frac{\overleftarrow{\delta}}{\delta\psi(x)}
   \right]
\notag\\
   &=4\left.\frac{\delta S^{(1)}}{\delta A_\mu (k)}\right|_{k=0}
   \int_p\,(-i)p_\mu
   \Tr\left\{
   \left[e^{ip(x-x')}+e^{ip(x'-x)}\right]
   \frac{\Slash{p}+im}{e^{-2p^2}+i\left(\Slash{p}+im\right)}
   \right\}
\notag\\
   &=4\left.\frac{\delta S^{(1)}}{\delta A_\mu (k)}\right|_{k=0}
   \int_p\,
   (-)p_\mu\Tr\left\{
   \left[e^{ip(x-x')}+e^{ip(x'-x)}\right]
   \frac{e^{-2p^2}+i\left(\Slash{p}+im\right)-e^{-2p^2}}
   {e^{-2p^2}+i\left(\Slash{p}+im\right)}
   \right\}
\notag\\
   &=4\left.\frac{\delta S^{(1)}}{\delta A_\mu (k)}\right|_{k=0}
   \int_p\,\left[e^{ip(x-x')}+e^{ip(x'-x)}\right]p_\mu
   \left\{-4+e^{-2p^2}\Tr\frac{1}{e^{-2p^2}+i\left(\Slash{p}+im\right)}
   \right\}.
\label{eq:(B6)}
\end{align}
Ignoring the derivative of the delta function, we obtain a vanishing
integral over~$p$:
\begin{equation}
   \stackrel{x'\to x}{\longrightarrow}
   8\left.\frac{\delta S^{(1)}}{\delta A_\mu(k)}\right|_{k=0}
   \int_p\,p_\mu e^{-2p^2}
   \Tr\frac{1}{e^{-2p^2}+i\left(\Slash{p}+im\right)}=0.
\label{eq:(B7)}
\end{equation}

\section{ERG for QED}
\label{sec:C}
In the ERG formalism, the Wilson action is constructed as
\begin{align}
   &e^{S_\Lambda[A_\mu,c,\Bar{c},\psi,\Bar{\psi}]}
\notag\\
   &\equiv\int\left[dA_\mu'dc'd\Bar{c}'d\psi'd\Bar{\psi}'\right]\,
   \exp\Biggl\{S\left[A_\mu',c',\Bar{c}',\psi',\Bar{\psi}'\right]
\notag\\
   &\qquad{}
   -\frac{\Lambda^2}{2}\int_k\,
   \left[A_\mu(k)-z_\Lambda e^{-k^2/\Lambda^2}A_\mu'(k)\right]
   \left[A_\mu(-k)-z_\Lambda e^{-k^2/\Lambda^2}A_\mu'(-k)\right]
\notag\\
   &\qquad{}
   -\Lambda^2\int_k\,
   \left[\Bar{c}(-k)-e^{-k^2/\Lambda^2}\Bar{c}'(-k)\right]
   \left[c(k)-e^{-k^2/\Lambda^2}c'(k)\right]
\notag\\
   &\qquad{}
   +i\Lambda\int_p\,
   \left[\Bar{\psi}(-p)-z_{F\Lambda}e^{-p^2/\Lambda^2}\Bar{\psi}'(-p)\right]
   \left[\psi(p)-z_{F\Lambda}e^{-p^2/\Lambda^2}\psi'(p)\right]
   \Biggr\},
\label{eq:(C1)}
\end{align}
where the electron fields are diffused according to the standard diffusion
equation. $z_\Lambda$ and~$z_{F\Lambda}$ here differ from those
in~Eq.~\eqref{eq:(3.1)}.

The Wilson action satisfies the ERG differential equation
\begin{align}
   &-\Lambda\frac{\partial}{\partial\Lambda}e^{S_\Lambda}
\notag\\
   &=\int_k\,
   \left[
   \left(2\frac{k^2}{\Lambda^2}-\gamma_\Lambda\right)A_\mu(k)
   \frac{\delta}{\delta A_\mu(k)}
   +\left(2\frac{k^2}{\Lambda^2}+1-\gamma_\Lambda\right) 
   \frac{1}{\Lambda^2}\frac{\delta^2}{\delta A_\mu(k)\delta A_\mu(-k)}\right]
   e^{S_\Lambda}
\notag\\
   &\qquad{}
   +\int_k\,
   \left[
   2\frac{k^2}{\Lambda^2}\Bar{c}(-k)
   \frac{\overrightarrow{\delta}}{\delta\Bar{c}(-k)}e^{S_\Lambda}
   +e^{S_\Lambda}\frac{\overleftarrow{\delta}}{\delta c(k)}
   2\frac{k^2}{\Lambda^2}c(k)
   -2\left(2\frac{k^2}{\Lambda^2}+1\right)
   \frac{1}{\Lambda^2}
   \frac{\overrightarrow{\delta}}{\delta\Bar{c}(-k)}e^{S_\Lambda}
   \frac{\overleftarrow{\delta}}{\delta c(k)}
   \right]
\notag\\
   &\qquad{}
   +\int_p\,
   \Biggl[
   e^{S_\Lambda}\frac{\overleftarrow{\delta}}{\delta\psi(p)}
   \left(2\frac{p^2}{\Lambda^2}-\gamma_{F\Lambda}\right)\psi(p)
   +\left(2\frac{p^2}{\Lambda^2}-\gamma_{F\Lambda}\right)\Bar{\psi}(-p)
   \cdot\frac{\overrightarrow{\delta}}{\delta\Bar{\psi}(-p)}e^{S_\Lambda}
\notag\\
   &\qquad\qquad\qquad{}
   -i\left(4\frac{p^2}{\Lambda^2}+1-2\gamma_{F\Lambda}\right) 
   \frac{1}{\Lambda}
   \Tr\frac{\overrightarrow{\delta}}{\delta\Bar{\psi}(-p)}
   e^{S_\Lambda}
   \frac{\overleftarrow{\delta}}{\delta\psi(p)}
   \Biggr].
\label{eq:(C2)}
\end{align}
This is the same as the first part of~Eq.~\eqref{eq:(3.11)}; the second part
proportional to~$e_\Lambda\equiv e\mu^{\epsilon/2}/z_\Lambda$ or~$e_\Lambda^2$
present in~Eq.~\eqref{eq:(3.11)} are missing here. The difference is due to the
simple diffusion equations we have adopted for the electron fields in ERG.

The BRST invariance of the original $S$ is inherited by the Wilson
action as
\begin{align}
   &\frac{1}{\xi_\Lambda}e^{-S_\Lambda}
   \int_k\,k_\mu
   \left[A_\mu(-k)+\frac{1}{\Lambda^2}\frac{\delta}{\delta A_\mu (k)}\right]
   \frac{\overrightarrow{\delta}}{\delta\Bar{c}(-k)}e^{S_\Lambda}
\notag\\
   &=\int_k\,e^{-k^2/\Lambda^2}k_\mu
   \left[c(k)\right]\frac{\delta}{\delta A_\mu(k)}S_\Lambda
\notag\\
   &\qquad{}
   -e_\Lambda e^{-S_\Lambda}\int_p\,
   e^{-p^2/\Lambda^2}
   \Tr\left[\int_k\,\left[c(k)\right]\left[\psi(p-k)\right] 
   e^{S_\Lambda}\right]
   \frac{\overleftarrow{\delta}}{\delta\psi(p)}
\notag\\
   &\qquad{}
   -e_\Lambda e^{-S_\Lambda}\int_p\,
   e^{-p^2/\Lambda^2}
   \frac{\overrightarrow{\delta}}{\delta\Bar{\psi}(-p)}
   \left[e^{S_\Lambda}\int_k\,\left[c(k)\right]\left[\Bar{\psi}(-p-k)\right]
   \right],
\label{eq:(C3)}
\end{align}
where we have defined the composite operators as
\begin{subequations}\label{eq:(C4)}
\begin{align}
   \left[c(k)\right]
   &\equiv e^{k^2/\Lambda^2}
   \left[
   c(k)+\frac{1}{\Lambda^2}\frac{\overrightarrow{\delta}}{\delta\Bar{c}(-k)}
   S_\Lambda\right],
\\
   \left[\psi(p)\right]
   &\equiv e^{p^2/\Lambda^2}
   \left[
   \psi(p)
   +\frac{i}{\Lambda}\frac{\overrightarrow{\delta}}{\delta\Bar{\psi}(-p)}
   S_\Lambda\right],
\\
   \left[\Bar{\psi}(-p)\right]
   &\equiv e^{p^2/\Lambda^2}
   \left[
   \Bar{\psi}(-p)
   +\frac{i}{\Lambda}S_\Lambda\frac{\overleftarrow{\delta}}{\delta\psi(p)}
   \right].
\end{align}
\end{subequations}
Since the ghost part of the action is given by
\begin{equation}
   S_{\Lambda,\text{ghost}}
   =-\int_k\,
   \Bar{c}(-k)\frac{k^2}{k^2/\Lambda^2+e^{-2k^2/\Lambda^2}}c(k),
\label{eq:(C5)}
\end{equation}
the BRST invariance reduces to the WT identity
\begin{align}
   &\frac{\xi e^{-2k^2}+k^2}{\xi e^{-k^2}}
   k_\mu\frac{\delta S_I}{\delta A_\mu(k)}
\notag\\
   &=ee^{-S}\int_p\,
   e^{-(p+k)^2+p^2}\Tr\left\{
   \left[\psi(p)+i\frac{\overrightarrow{\delta}}{\delta\Bar{\psi}(-p)}\right]
   e^S\right\}\frac{\overleftarrow{\delta}}{\delta\psi(p+k)}
\notag\\
   &\qquad{}
   -ee^{-S}\int_p\,
   e^{-p^2+(p+k)^2}\Tr\frac{\overrightarrow{\delta}}{\delta\Bar{\psi}(-p)}
   \left\{
   e^S
   \left[
   \Bar{\psi}(-p-k)+i\frac{\overleftarrow{\delta}}{\delta\psi(p+k)}
   \right]\right\}
\label{eq:(C6)}
\end{align}
in the dimensionless notation. Because of the mismatch of the exponential
cutoff functions, the WT identity is non-linear in~$S$, not as simple as our WT
identity~\eqref{eq:(5.13)}.



%\bibliographystyle{ptephy}
%\bibliography{sample}
%
% once the .bbl file has been generated then place the text in your article.

%% \vspace{0.2cm}
%% \noindent
%% For references,  note how to include DOI information from examples below. 

%This is added by T. Yoneya (editor-in-chief) on 2020/07/09.

\let\doi\relax

%without this code before the command "\begin{thebibliography}{}" , an error will be %flagged. When the bibliography is provided as separate .bib file, then this code %should be placed above the commands "\bibliographystyle{}" and "\bibliography{}" %inside the main TeX file. 

%% \begin{thebibliography}{9}

%% \bibitem{1}
%% J. P.~Blaizot, and E.~Iancu, Phys. Rep. {\bf 359}, 355 (2002).
%% \doi{https://doi.org/10.1016/S0370-1573(01)00061-8}

%% \bibitem{2}
%% M.~Gyulassy, and L.~McLerran, Nucl.\ Phys.\  A {\bf 750}, 30 (2005). \\ \doi{https://doi.org/10.1016/j.nuclphysa.2004.10.034}

%% \bibitem{3}
%% S.~Aoki et al. [JLQCD Collaboration], Phys. Rev. D 72, 054510 (2005). \\
%% \doi{https://doi.org/10.1103/PhysRevD.72.05451}

%% \bibitem{4}
%% S.~Alekhin, A.~Djouadi, and S.~Moch, Phys. Lett. B 716, 214 (2012) [arXiv:1207.0980 [hep-ph]]. \doi{https://doi.org/10.1016/j.physletb.2012.08.024}

%% \end{thebibliography}

\bibliographystyle{ptephy}
\bibliography{manifestly}

\end{document}